\documentclass[a4paper,11pt]{article}
\pdfoutput=1 

\usepackage{jheppub}
\usepackage{float}
\usepackage{ulem}
\usepackage{mathrsfs}
\usepackage{natbib}
\usepackage{tikz}
\usetikzlibrary{calc,cd,backgrounds, 
  snakes}
\usetikzlibrary{decorations.pathmorphing}
\usepackage{setspace}
\usepackage{amsfonts}
\usepackage{amssymb}
\usepackage{amsmath}
\usepackage{esint}                   
\usepackage{epsfig}
\usepackage{latexsym}
\usepackage{color}
\usepackage{graphicx}
\usepackage{hyperref}
\usepackage{nicefrac}
\usepackage[latin1]{inputenc}
\usepackage{pstricks}
\usepackage{slashed}
\usepackage{multirow}
\usepackage{subcaption}

\def\be{\begin{equation}}
\def\ee{\end{equation}}
\def\ba{\begin{eqnarray}}
\def\ea{\end{eqnarray}}

\def\r{\rho}
\def\a{\alpha}

\def\b{\beta}

\def\th{\theta}

\def\m{\mu}
\def\n{\nu}

\def\Om{\Omega}
\def\l{\lambda}

\def\s{\sigma}

\def\cN{{\cal N}}

\def\IR{\relax{\rm I\kern-.18em R}}

\def\inv{^{\raise.0ex\hbox{${\scriptscriptstyle -}$}\kern-.05em 1}}

\begin{document}
\hfill HU-EP-25/19

\title{Universal Observables, SUSY RG-Flows and Holography}

\author[a]{Dimitrios Chatzis, }
\author[a]{Madison Hammond, }
\author[b]{Georgios Itsios, }
\author[a]{Carlos Nunez  }
\author[c]{ and 
Dimitrios Zoakos}
\affiliation[a]{Department of Physics, Swansea University,\\
Swansea SA2 8PP, United Kingdom}
\affiliation[b]{Institut f\"{u}r Physik, Humboldt-Universit\"{a}t zu Berlin,\\
IRIS Geb\"{a}ude, Zum Gro{\ss}en Windkanal 2, 12489 Berlin, Germany}
\affiliation[c]{Department of Physics, University of Patras, 26504 Patras, Greece}

\emailAdd{dchatzis@proton.me}
\emailAdd{m.hammond.2412736@swansea.ac.uk}
\emailAdd{georgios.itsios@physik.hu-berlin.de}
\emailAdd{c.nunez@swansea.ac.uk}
\emailAdd{dzoakos@upatras.gr}
\abstract{
We construct and analyse infinite classes of regular supergravity backgrounds dual to four-dimensional
superconformal field theories (SCFTs) compactified on a circle with a supersymmetry-preserving twist. These flows lead to three-dimensional gapped QFTs preserving four supercharges. The solutions arise in Type IIB, Type IIA, and eleven-dimensional supergravity, and generalise known constructions by incorporating deformations that avoid typical singularities associated with the holographic description of the Coulomb branch of the CFT. We examine several observables: Wilson loops, holographic central charges, and complexity. We show they exhibit a universal factorisation, with each observable decomposing into a UV-CFT contribution times a flow-dependent factor. We also explore the parameter regimes where higher-curvature corrections become relevant, affecting the physical interpretation of certain observables. Our findings provide new insights into universal features of holographic RG flows and resolve a puzzle related to complexity in these systems.
}

\maketitle

\flushbottom


\section{Introduction}
The AdS/CFT correspondence and its refinements \cite{Maldacena:1997re,Gubser:1998bc,Witten:1998qj} have been naturally extended to  use  holographic techniques for the study of strongly coupled, non-conformal field theories.

There are two well-established approaches to constructing non-singular holographic duals of confining quantum field theories (QFTs). The first involves wrapped branes--see for example \cite{Witten:1998zw,Maldacena:2000yy,Atiyah:2000zz,Edelstein:2001pu}. The second deforms the D3-brane solution on the conifold by introducing fractional branes, establishing a connection with quiver gauge theories \cite{Klebanov:1998hh,Klebanov:2000nc,Klebanov:2000hb,Gubser:2004qj}. Both approaches have been extensively explored and generalized. Moreover, it is possible to relate these two constructions \cite{Maldacena:2009mw,Gaillard:2010qg, Caceres:2011zn}.

A common limitation of the above systems is that, although the infrared (IR) dynamics of the dual QFTs is well understood--thanks to reliable string dual descriptions--the ultraviolet (UV) behaviour remains problematic. Wrapped brane constructions often imply a higher dimensional UV completion, which may not correspond to a conventional field theory. Meanwhile, quiver based models do not precisely flow to a conformal fixed point. In both scenarios, the number of degrees of freedom from a four-dimensional perspective grows unbounded with energy. Additionally, incorporating fundamental matter (degrees of freedom in the fundamental representation of the gauge group) is technically challenging, see \cite{Casero:2006pt,Nunez:2010sf,Benini:2006hh,Benini:2007gx,Bigazzi:2008qq,Bigazzi:2014qsa}.

In this work, we make progress in addressing these challenges. We construct infinite classes of string theory backgrounds dual to four-dimensional superconformal field theories (SCFTs) that flow in the IR to $(2+1)$-dimensional ${\cal N}=2$ gapped QFTs. We discuss various observable quantities and elaborate on the possible corrections to the supergravity solutions. Further technical and conceptual details about our construction can be found in the companion paper \cite{CHINZ2}.

The new backgrounds correspond to families of 4d SCFTs deformed by vacuum expectation values (VEVs) and compactification on a circle. This compactification is a twisting procedure. The resulting low-energy theories are three-dimensional, preserve four supercharges, and are defined on vacua that give a VEV to a global current and a dimension-two operator. These field theories may be Lagrangian or non-Lagrangian, single-node or linear quivers with flavours. From a field-theoretic perspective, the SCFT's R-symmetry is exploited to preserve a subset of the original supersymmetry, which would otherwise be broken by compactification on the circle $S^1$. In the case of ${\cal N}=4$ SYM, this mechanism has been analyzed in detail by Kumar and Stuardo \cite{Kumar:2024pcz}, and initially developed in the work of Cassani and Komargodski \cite{Cassani:2021fyv}. The main technical tools used in our construction are drawn from \cite{Anabalon:2021tua,Anabalon:2022aig,Gauntlett:2007sm,Anabalon:2024che}.

\subsection{General Idea and Outline}\label{QFTapproach}
When placing a supersymmetric QFT on a background of the form $R^{1,d} \times S^1_\phi$, one must define boundary conditions for fermions, scalars, and gauge fields along the compact circle $S^1_\phi$. Typically, supersymmetry is broken in this setting, since scalars and gauge fields satisfy periodic boundary conditions, whereas fermions are anti-periodic.

However, supersymmetry can be partially preserved by mixing the QFT's R-symmetry with the component of the Lorentz group associated with translations along $S^1_\phi$. This idea is explained clearly for ${\cal N}=4$ SYM in \cite{Kumar:2024pcz}, and for generic SCFTs in \cite{Cassani:2021fyv}.

Indeed, turning on a constant background R-symmetry gauge field $\mathcal{A} = \mathcal{A}_\phi  \mathrm{d}\phi$ modifies the covariant derivative in a way that allows massless fermions to survive. These pair-up with gauge fields and scalars  to form massless supermultiplets. Similarly, massive fermionic and scalar modes organize into multiplets, preserving four supercharges. The result is an ${\cal N}=2$ supersymmetric  theory in three dimensions, supplemented by massive excitations. Although the background gauge field is constant, it has non-trivial holonomy, which has dynamical consequences.


The holographic dual of this supersymmetry-preserving twist has been constructed by Anabal\'on and Ross \cite{Anabalon:2021tua}. For additional studies employing similar mechanisms, see \cite{Nunez:2023nnl, Anabalon:2024che, Anabalon:2024qhf, Fatemiabhari:2024aua, Nunez:2023xgl, Anabalon:2023lnk, Anabalon:2022aig, Macpherson:2024qfi}.

One aim of this work is to holographically realize the above partial supersymmetry-preserving mechanism for broad families of 4d SCFTs compactified on $S^1$. Our examples include single-node CFTs, an infinite family of non-lagrangian CFTs and an infinite family of linear conformal quivers. We explicitly construct supergravity backgrounds that encode the geometric realization of this mechanism and proceed to compute a range of field-theoretic observables. 

Several of the observables we compute exhibit a universal behaviour. In certain parameter regimes, our supergravity backgrounds require corrections. We analyze how these corrections affect key observables.

We focus on the holographic dual of a deformation of ${\cal N}=4$
SYM into its Coulomb branch. The original solution in \cite{Freedman:1999gp, Freedman:1999gk} features a naked singularity, but we work with a regularised version obtained via the twisted compactification mentioned earlier. This desingularised solution was constructed in gauged supergravity in \cite{Anabalon:2024che}.

Our supergravity backgrounds depend on several parameters--one of which controls higher-order invariants such as 
$R_{ab}^2$ and $R_{abcd}^2$. Even when the geometry is smooth, these invariants can become large. We study their impact on observables like the Wilson loop.

Importantly, many observables show universal behaviour. By this we mean that their values factorise into a term coming from the UV-CFT and another coming from the RG flow. This follows from a conjecture by Gauntlett and Varela \cite{Gauntlett:2007ma}, later proven in \cite{Cassani:2019vcl}. We leverage this result to resolve a puzzle related to quantum complexity in our QFTs.
\\
\\
The paper is organized as follows:
Section \ref{SUGRA_backgrounds} presents three classes of supergravity backgrounds (class I was introduced in \cite{Anabalon:2024che}). We discuss their singularity structure and corresponding Page charges.
Section \ref{sectionQFT} explores the dual field theory aspects of these backgrounds. We compute observables such as Wilson loops, holographic central charges, and complexity, all of which exhibit the universal structure mentioned above. These results reveal that even in regular backgrounds, corrections--controlled by specific parameters--can significantly influence the Physics. We also resolve a conceptual issue concerning complexity.
We conclude in Section \ref{conclusionsection}.

\section{Supergravity backgrounds}\label{SUGRA_backgrounds}
In this section we present three supergravity backgrounds. These are solutions of the equations of motion of Type IIB (for background I), eleven dimensional supergravity (backgrounds II) and Type IIA (backgrounds III). These all share a common five dimensional sub-manifold, that was written by Anabal\'on, Nastase and Oyarzo \cite{Anabalon:2024che}. The equations of motion have been checked using Mathematica. The construction of each of these backgrounds, the presence of SUSY and other details are discussed in the companion work \cite{CHINZ2}. {We give a quick summary of the construction in Section \ref{rough-view}}. In Section \ref{sectionQFT} we study some aspects of the dual field theories and observables.  Below, we write the backgrounds.
\subsection{Background of class I}\label{background2b}
We start writing the first of our backgrounds. This is a solution to the equations of motion of Type IIB supergravity with metric and five form (all the other fields vanish). We choose to describe it in
terms of the coordinates $[t,w,z,\phi, r, \theta,\psi,\phi_1,\phi_2,\phi_3]$. We use a vielbein basis,
\begin{equation}\label{S5-1}
 \begin{aligned}
  & e^0 = \frac{\sqrt{\zeta}}{L} \, r \, dt \, , \qquad
     e^1 = \frac{\sqrt{\zeta}}{L} \, r \, dw \, , \qquad
     e^2 = \frac{\sqrt{\zeta}}{L} \, r \, dz \, , \qquad
     e^3 = \sqrt{\zeta \, F} \, r \, d\phi \, ,
 \\[5pt]
  & e^4 = \sqrt{\frac{\zeta}{F}} \frac{dr}{\l^3 r} \, , \qquad
     e^5 = L \sqrt{\zeta} \, d\th \, , \qquad
     e^6 = L \frac{\cos\th}{\sqrt{\zeta}} \, d\psi \, ,
  \\[5pt]
  & e^7 = L \frac{\cos\th \, \sin\psi}{\sqrt{\zeta}} \, \Big( d\phi_1 + \frac{A_1}{L} \Big) \, , \qquad
     e^8 = L \frac{\cos\th \, \cos\psi}{\sqrt{\zeta}} \, \Big( d\phi_2 + \frac{A_2}{L} \Big) \, ,
 \\[5pt]
 & e^9 = L \frac{\l^3 \sin\th}{\sqrt{\zeta}} \, \Big( d\phi_3 + \frac{A_3}{L} \Big) \, .
 \end{aligned}
\end{equation}
The metric is $ds^2= \eta_{ab}e^a e^a$ (being $\eta_{ab}$ the ten dimensional Minkowski metric),
\begin{eqnarray}
& & ds^2= \frac{\zeta(r,\theta)}{L^2}\left[r^2\left( -dt^2+ dw^2+dz^2 + L^2 F(r) d\phi^2\right) +\frac{L^2 dr^2}{F(r) r^2\lambda^3(r)} + L^4 d\theta^2 \right]\label{metricaS5}\\
& & +\frac{L^2}{\zeta(r,\theta)}\Bigg[\cos^2\theta d\psi^2+\cos^2\theta\sin^2\psi \left(d\phi_1 +\frac{A_1}{L}\right)^2 +\cos^2\theta\cos^2\psi \left(d\phi_2 +\frac{A_2}{L}\right)^2 \nonumber\\ & & +\lambda^6(r) \sin^2\theta \left( d\phi_3 +\frac{A_3}{L}\right)\Bigg].\nonumber
\end{eqnarray}
In the expressions above $L$ is a constant (the 'AdS-radius'). The functions $\zeta=\zeta(r,\theta)$, $\l=\lambda(r)$ and $F=F(r)$ are
\begin{equation}
 \begin{aligned}
  & \zeta(r , \th) = \sqrt{1 + \varepsilon \frac{\ell^2}{r^2} \cos^2\th} \, , \qquad 
 \l(r) = \Big( 1 + \varepsilon \frac{\ell^2}{r^2}  \Big)^{\frac{1}{6}} \, ,
 \\[5pt] 
  & F(r) = \frac{1}{L^2} - \frac{\varepsilon \ell^2 L^2}{ r^4} \Big( q^2_1 - \frac{q^2_2}{\l^6} \Big) \, .
 \end{aligned}\label{lionel1}
\end{equation}
Moreover, $ \, \ell , \, q_1 , \, q_2 $ are constants and $\varepsilon=\pm 1$ is a sign.
%
%
Finally, $A_i \, (i = 1,2,3)$ is a set of one-forms such that
\begin{eqnarray}\label{gauge_fields}
 & & A_1 = A_2 = A_{1\phi} d\phi=\varepsilon \, q_1 \, \ell^2 \, L \, \frac{r^2 - r^2_\star}{r^2_\star \, r^2} \, d\phi \, ,\\ 
& & A_3 = A_{3\phi}d\phi=\varepsilon \, q_2 \, \ell^2 \, L \, \frac{r^2 - r^2_\star}{ (r^2_\star + \varepsilon \, \ell^2) \, (r^2 + \varepsilon \, \ell^2) } \, d\phi \, .
\end{eqnarray}
The constant $r_\star$ is the value of the largest root of $F(r)$. In what follows, we set $q_1=q_2$, in which case, the background preserves four Poincare SUSYs.
The self-dual RR five-form is
\begin{equation}
 F_5 = (1 + \star) G_5 \, .
\end{equation}
In term of the vielbeins of eq.(\ref{S5-1}) 
%

\begin{equation}\label{typeIIB_flux_vielbein}
    \begin{split}
        G_5 &=\frac{2\lambda^3(1+\zeta^2)}{L\zeta^{5/2}}e^{01234}-\frac{\varepsilon\ell^2\sqrt{F}\sin(2\theta)}{r^2\zeta^{5/2}}e^{01235}-\frac{\sin\theta\sin\psi A^{\prime}_{1\phi}}{\zeta^{3/2}}e^{01257}\\
        &-\frac{\sin\theta\cos\psi A^{\prime}_{1\phi}}{\zeta^{3/2}}e^{01258}+\frac{\cos\psi A^{\prime}_{1\phi}}{\sqrt{\zeta}}e^{01267}-\frac{\sin\psi A^{\prime}_{1\phi}}{\sqrt{\zeta}}e^{01268}+\frac{\cos\theta A^{\prime}_{3\phi}\lambda^9}{\zeta^{3/2}}e^{01259},
    \end{split}
\end{equation}
where $e^{abcpq}=e^a\wedge e^b\wedge e^c\wedge e^{p}\wedge e^q$. Also, we denoted with a prime the derivative  with respect to the $r$-coordinate. As a side note, the units of the coordinates and parameters are: $\ell,L$ have units of length while $q_1,q_2$ have units of inverse length. The coordinates $[t,w,z,\phi,r]$ have length dimension, whilst $[\theta,\psi,\phi_1,\phi_2,\phi_3]$ are dimensionless. $F(r)$ has units of inverse-length-squared and the one forms $A_1, A_2,A_3$ have units of length.
\\
Let us now connect the solution in eqs.(\ref{S5-1})-(\ref{typeIIB_flux_vielbein}) with other related backgrounds.
\\
\underline{\bf Connecting with Anabal\'on-Ross and AdS$_5\times S^5$ }
\\
In the solution of eqs.(\ref{S5-1})-(\ref{typeIIB_flux_vielbein}) , we have parameters $q_1,q_2,\ell, L$ and the choice of sign for $\varepsilon=\pm 1$. The case that mostly interests us is $q_1=q_2$ (that preserves SUSY). Hence, let  us take $q_1=q_2=Q$ and define
\begin{equation}
q_1=q_2=Q=\frac{\hat{q}}{\ell^2}.  \label{zizu}  
\end{equation}
Note that $\hat{q}$ has units of length. Consider the limit $\ell\to 0$, keeping $\hat{q}$ fixed. In this case we find
\begin{equation}
\zeta(r,\theta)=1 \, , \quad 
A_1\!=\!A_2\!=\!A_3= \varepsilon \, \hat{q} \, L \left( \frac{1}{r_{\star}^2} -\frac{1}{r^2}\right)d\phi \, , \quad 
F(r)= \frac{1}{L^2}\left(1-\frac{L^4 \hat{q}^2}{r^6}\right).    
\label{zivu}
\end{equation}
The solution now contains the parameters $\hat{q}, L$. The choice $\varepsilon=\pm 1$ is not physically relevant, as it is just an orientation for the $\phi$-direction. This is the solution found by Anabal\'on and Ross in \cite{Anabalon:2021tua} and further studied in \cite{Anabalon:2022aig}.  
We can further truncate by taking $\hat{q}=0$, finding AdS$_5\times S^5$. In this way we connect these three solutions by flowing in the parameter-space of the background in eqs.(\ref{S5-1})-(\ref{typeIIB_flux_vielbein}).
\\
Let us now discuss the regularity conditions or singular behaviour of the above solution of Type IIB.


\subsubsection{Singularity structure}\label{singularity_section}

To study the presence or absence of singularities we need a better understanding of the  functions $\lambda(r)$ and $F(r)$ defining this background. As above, we focus on the supersymmetric case  $q_1=q_2=Q$. The function $F(r)$ can be written as\footnote{We use $F(r_{\star})=0$ to eliminate $Q$. In this re-writing of $F$, we make clear that $r_\star$ is a  root.}
\begin{equation}
\begin{split}
    F(r)=(r^2-r_{\star}^2)\left(\frac{r^4+(r_{\star}^2+\varepsilon\ell^2)r^2+r_{\star}^2(r_{\star}^2+\varepsilon\ell^2)}{L^2r^4(r^2+\varepsilon\ell^2)}\right),
    \end{split}
\end{equation}
where $r\in [r_\star,\infty)$. In the $\mathrm{UV}$ limit ($r\to\infty$), we have $F(r)\to\frac{1}{L^2}$ and $\lambda(r)\to 1$. The metric asymptotes to $\mathrm{AdS}_5\times S^5$. It is convenient to express things in terms of a dimensionless variable $\xi$ and a parameter $\hat{\nu}$ defined as,
\begin{equation}
    \xi=\frac{r}{r_\star}, ~~~~\hat{\nu}=\varepsilon\frac{\ell^2}{r_\star^2},
\end{equation}
The functions defining the background read,
\begin{equation}\label{lambdaandFwithnu}
    \begin{split}
        &\lambda^6(\xi)=\frac{\xi^2+\hat{\nu}}{\xi^2},~~A_{1\phi}=\hat{\nu} Q L (1-\frac{1}{\xi^2}),~~A_{3\phi}= \frac{\hat{\nu} Q L (\xi^2-1)}{(1+\hat{\nu})(\xi^2+\hat{\nu})},~~\zeta(\xi,\theta)=\sqrt{1+\frac{\hat{\nu}}{\xi^2}\cos^2\theta}\\
        &F(\xi)=\frac{(\xi^2-1)\left[\xi^4+(1+\hat{\nu})\xi^2+1+\hat{\nu}\right]}{L^2\xi^4(\hat{\nu}+\xi^2)} = \frac{(\xi^2-1)\left(\xi^2-\xi^2_+ \right)\left(\xi^2 - \xi^2_-\right)}{L^2\xi^4(\hat{\nu}+\xi^2)},\quad \xi\geq 1.
    \end{split}
\end{equation}
One can solve the sixth order polynomial equation $F(r_\star)=0$ directly to obtain an expression of $r_\star$ in terms of the rest of the parameters:
\begin{equation}\label{rstar_explicit}
    \begin{split}
        &r_\star=\frac{1}{\sqrt{6}}\sqrt{2^{2/3}\ell^{4/3}\Lambda + 2\ell^{2}\varepsilon\left(-1+\frac{2^{1/3}\ell^{2/3}\varepsilon}{\Lambda}\right)},\\[5pt]
        &\Lambda:=\left[ -2\varepsilon\ell^2+3Q\left(9L^4Q+L^2\sqrt{81L^4Q^2-12\ell^2\varepsilon}\right) \right]^{1/3},
    \end{split}
\end{equation}
 Substituting this in $\hat{\nu}$, we find that it is positive for every Q when $\varepsilon=+1$, while $\hat{\nu}>-1$ when $\varepsilon=-1$, where the value -1 occurs when $Q=0$.

\begin{figure}[h!]
    \centering
    \begin{subfigure}[b]{0.45\textwidth}
        \includegraphics[width=\textwidth]{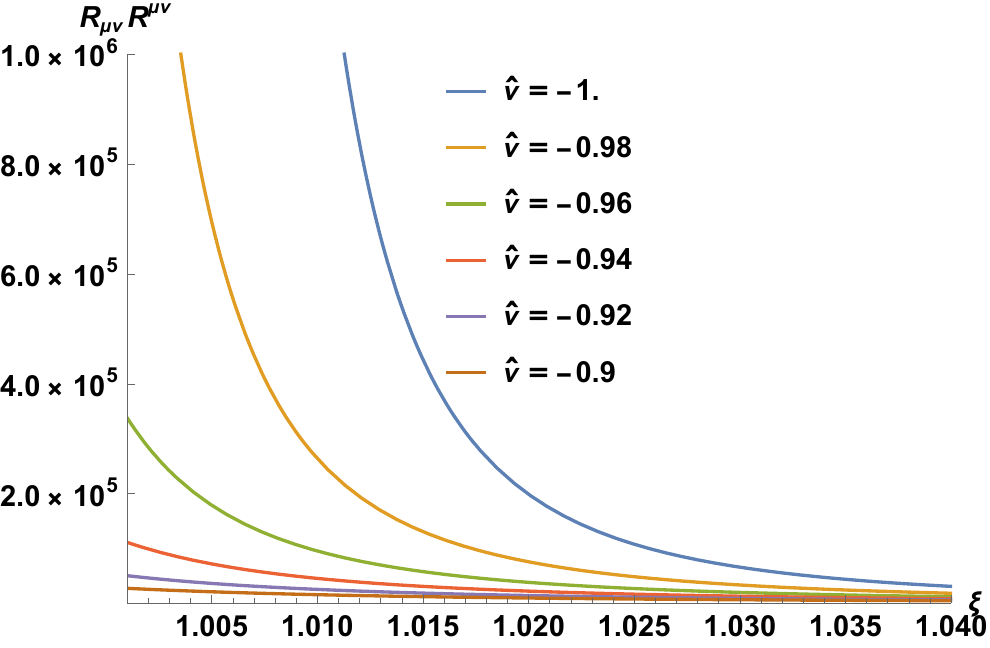}
        \caption{The invariant $R_{\m\n} \, R^{\m\n}$ at $\th = 0$}
    \end{subfigure}
    ~ 
    \begin{subfigure}[b]{0.45\textwidth}
        \includegraphics[width=\textwidth]{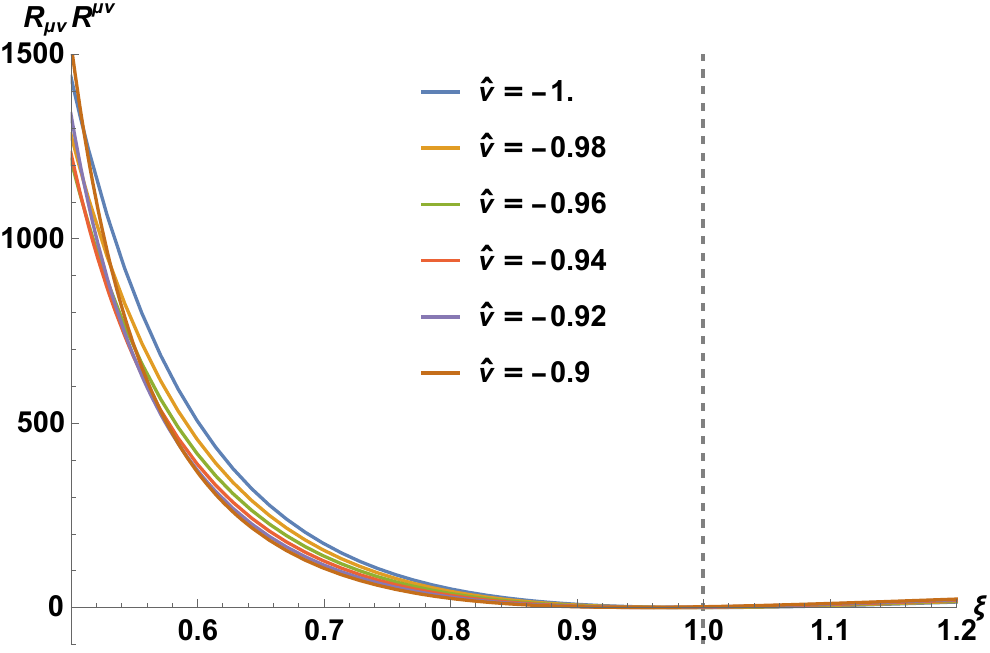}
        \caption{The invariant $R_{\m\n} \, R^{\m\n}$ at $\th = \frac{\pi}{2}$}
    \end{subfigure}
    \captionsetup{width=.75\textwidth,justification=centering}
    \caption{The Ricci tensor squared for $\th = 0$ and  $\th= \frac{\pi}{2}$ with
    \\
    $\hat{\n} = -1, -0.98, -0.96, -0.94, -0.92, -0.9$. In figure (b) the end of the space is shown by the dashed line.}\label{figuraricci}
\end{figure}

\begin{figure}[h!]
    \centering
    \begin{subfigure}[b]{0.45\textwidth}
        \includegraphics[width=\textwidth]{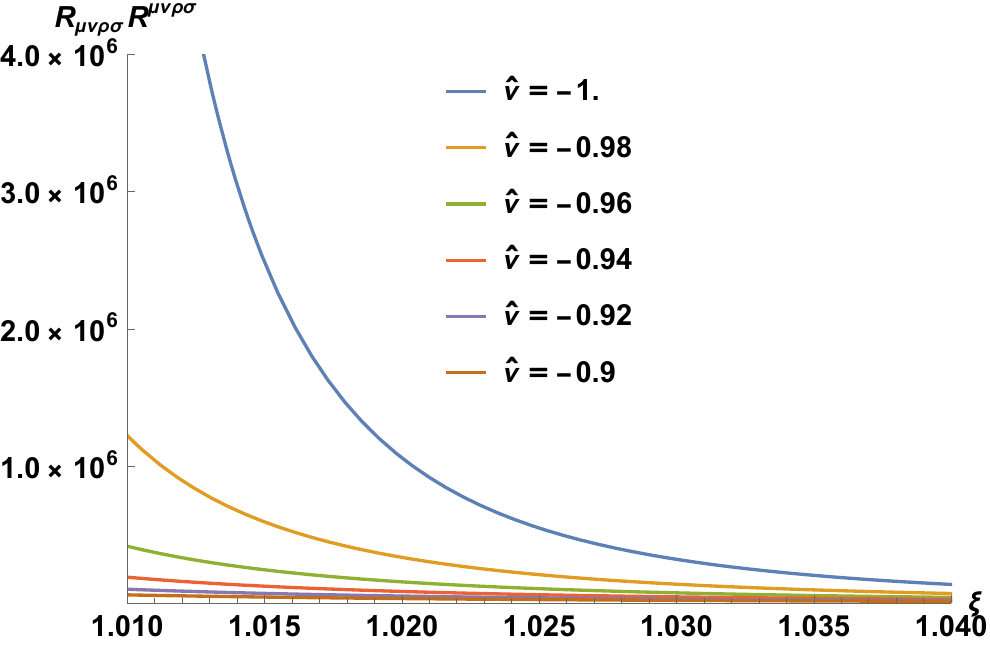}
        \caption{The invariant $R_{\m\n\r\s} \, R^{\m\n\r\s}$ in terms of $\xi$, at $\th = 0$}
    \end{subfigure}
    ~ 
    \begin{subfigure}[b]{0.45\textwidth}
        \includegraphics[width=\textwidth]{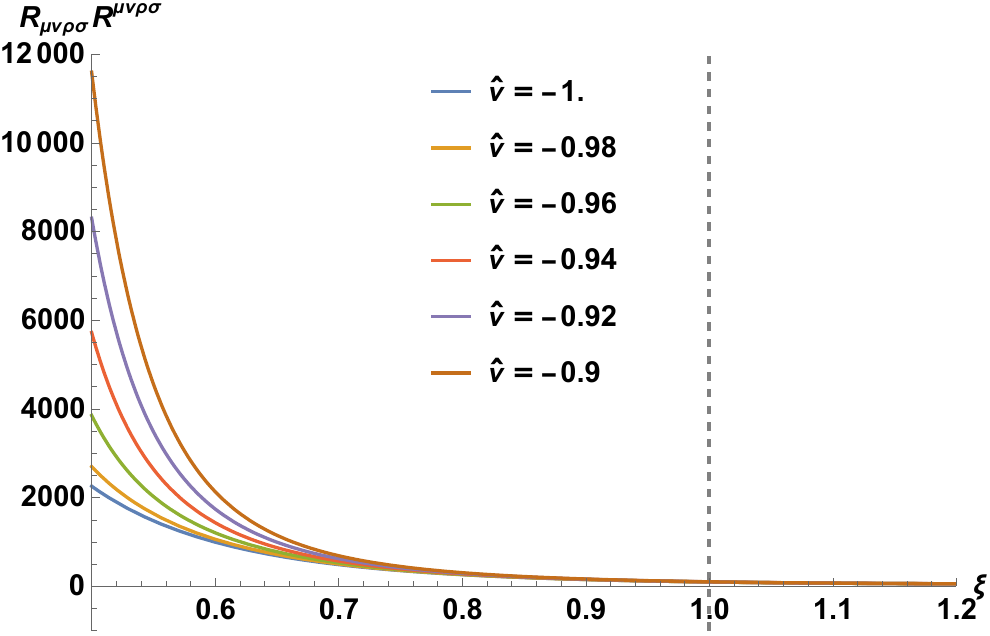}
        \caption{The invariant $R_{\m\n\r\s} \, R^{\m\n\r\s}$ as a function of $\xi$  at $\th = \frac{\pi}{2}$}
    \end{subfigure}
    \captionsetup{width=.75\textwidth,justification=centering}
    \caption{The Riemann tensor squared for $\th = 0$  and $\theta=\frac{\pi}{2}$ with
    \\
    $\hat{\n} = -1, -0.98, -0.96, -0.94, -0.92, -0.9$. In figure (b) the end of the space is shown by the dashed line.}
    \label{figurariemann}
\end{figure}

\begin{figure}[h!]
    \centering
    \begin{subfigure}[b]{0.45\textwidth}
        \includegraphics[width=\textwidth]{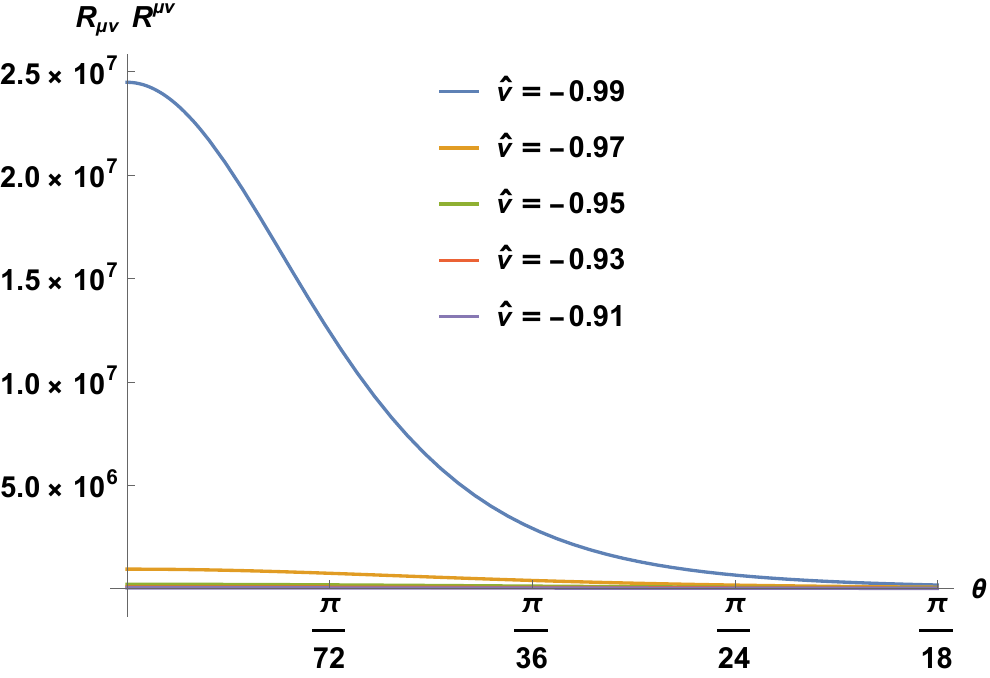}
        \caption{The invariant $R_{\m\n} \, R^{\m\n}$ at $\xi=1$ for different values of $\theta$ and parameter $\hat{\nu}$.}
    \end{subfigure}
    ~ 
    \begin{subfigure}[b]{0.45\textwidth}
        \includegraphics[width=\textwidth]{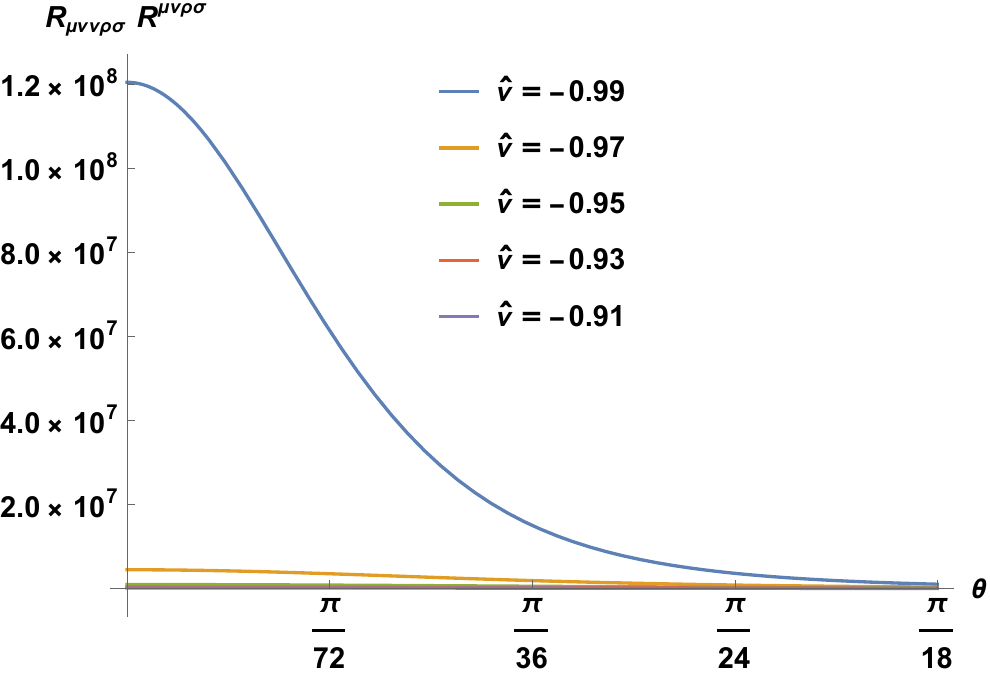}
        \caption{The invariant $R_{\m\n\rho\sigma} \, R^{\m\n \rho\sigma}$ at $\xi = 1$ for different values of $\theta$ and different $\hat{\nu}$.}
    \end{subfigure}
    \captionsetup{width=.75\textwidth,justification=centering}
    \caption{The Ricci tensor squared  and the Riemann tensor squared starting from $\theta=0$, for $\xi=1$ and varying $\hat\nu$. }
    \label{riccisquared}
\end{figure}

Let us study the root structure of $F(\xi)$ in eq. (\ref{lambdaandFwithnu}) for different values of the parameter $\hat{\nu}\geq-1$. The roots of the fourth order polynomial in the numerator of $F$ are:
\begin{equation}
    \xi_{\pm}^2=\frac{-(1+\hat{\nu})\pm\sqrt{\big.(\hat{\nu}+1)(\hat{\nu}-3)}}{2}.
\end{equation}
We make the following observations:
\begin{itemize}
    \item  For $\hat{\nu}\in(-1,\infty)$ there are no real roots for $\xi_\pm$. The space ends at $\xi=1$, where $F(\xi)=0$.
    \item The value $\hat{\nu}=-1$, which belongs to the $\varepsilon=-1$ branch of the supergravity solution, gives $\xi_\pm=0$,  $F(\xi)=L^{-2}$,  
    and $\lambda(\xi=1)$ vanishes. We claim that in this case ($\hat{\nu}=-1$), the background has a singularity at $\xi=1$  (or, at $r=r_\star\equiv\ell$).
    %
\end{itemize}
To justify the last statement, we compute the invariants $R_{ab}^2$ and $R_{abcd}^2$. For $\hat{\nu}=-1$ (the lower of the allowed values of the parameter) we find,
\begin{equation}
R_{\m\n} R^{\m\n} = - \frac{10}{L^4} \frac{\big( \cos^2\th - 4 \xi^2 + 4 \xi^4 \big)^2}{\xi^2 \big( \cos^2\th - \xi^2 \big)^3}
\end{equation}
and
\begin{eqnarray}
 R_{\m\n\r\s} R^{\m\n\r\s} & =& - \frac{51 - 352 \xi^2 + 1568 \xi^4 - 1280 \xi^6 + 640 \xi^8}{8 L^4 \xi^2 \big( \cos^2\th - \xi^2 \big)^3}\nonumber\\
& &  - \frac{4 \big( 17 - 88 \xi^2 + 40 \xi^4 \big) \cos(2 \th) + 17 \cos(4 \th)}{8 L^4 \xi^2 \big( \cos^2\th - \xi^2 \big)^3} \, .
\end{eqnarray}
Both expressions indicate that the geometry is singular along the curve 
\begin{equation}
 \cos^2\th - \xi^2 = 0 \, .
\end{equation}
In other words, for $\hat{\nu}=-1$, the end of the space $\xi=1$ is singular when we approach it in the direction $\theta=0$. In Figures \ref{figuraricci} and \ref{figurariemann} we illustrate the behaviour of the two invariants for various values of $\hat{\n}$ (around $\hat{\nu}\approx -1$) for two values of the angle $\th = 0$ and $\th = \frac{\pi}{2}$. Note that the space ends at $\xi=1$. In particular, note that even for $\hat{\nu}=-1$ there is no singularity approaching the end of the space in a direction $\theta>0$. 
\\
\underline{\bf An observation }
\\
Consider the case in which the space is {\it non-singular}  (namely $-1<\hat{\nu}<\infty$). A natural question is if there is any effect close to the end of the space (at $\xi=1$). In particular if for values of the parameter $\hat{\nu}$ close to the edge of allowed values $\hat{\nu}\sim -1^+$ we find any reason for the supergravity approximation to be invalid.
Since studying this analytically is not straightforward, we plot $R_{ab}R^{ab}$ and $R_{abcd}R^{abcd}$ for the point $\xi=1$ and for different values of $\theta$ close to $\theta=0$, see also \cite{Brandhuber:1999jr}. We also choose the parameter $\hat\nu\to -1^+$. Figure \ref{riccisquared} displays this analysis.
%

%
%

In conclusion, as we approach the end of the space $\xi=1$ in the direction $\theta=0$, with values of $\hat{\nu}\approx-1$ we encounter that the supergravity approximation is not necessarily valid and corrections to the background should be considered.
This has consequences for the dual QFT  observables computed using the holographic background, as we discuss in Section \ref{wilsonsection}.

%
%
To close the  analysis of singularities, let us discuss the absence of conical singularities. It is important to study the metric of the $(r,\phi)$ subspace near $r=r_\star$ and find the periodicity of ${\phi}$ that avoids conical singularities. We use the following expansions for the functions:
\begin{eqnarray}
& &\lambda(\xi)\approx (1+\hat{\nu})^{1/6}+\mathcal{O}(\xi-1), ~~~
F(\xi)\approx \frac{2(2\hat{\nu}+3)(\xi-1)}{L^2(1+\hat{\nu})}+\mathcal{O}((\xi-1)^2),\\
& & A_{1\phi}\approx 2 L Q\hat{\nu}(\xi-1)+ \mathcal{O}((\xi-1)^2), ~~ A_{3\phi}\approx \frac{2{L Q \hat{\nu}}}{(1+\hat{\nu})^2}(\xi-1)+ \mathcal{O}((\xi-1)^2),\nonumber\\
& & \zeta(\xi,\theta)\approx \sqrt{1+\hat{\nu}\cos^2\theta}+\mathcal{O}(\xi-1).\nonumber
\end{eqnarray}
We focus on the $(r,\phi)$ subspace. The result of expanding the metric reads, 
\begin{equation}
    \mathrm{d}s^2_{r,\phi} \approx \sqrt{1+\hat{\nu}\cos^2\theta} \left[ \frac{L^2}{(4\hat{\nu}+6)(\xi-1)}\mathrm{d}\xi^2 + \frac{r_\star^2 (\xi -1)(4\hat{\nu} + 6)}{L^2 (1+\hat{\nu})}\mathrm{d}\phi^2 \right].
\end{equation}
Defining a  new radial coordinate as
$
    \rho = 2L \sqrt{\frac{\xi - 1}{4\hat{\nu}+6}},$
we obtain
\begin{equation}
     \mathrm{d}s^2_{r,\phi} \approx \sqrt{1+\hat{\nu}\cos^2\theta} \left[ \mathrm{d}\rho^2 + \frac{ r_\star^2 (4\hat{\nu} + 6)^2}{4L^4 (1+\hat{\nu})} \; \rho^2 \;\mathrm{d}\phi^2 \right].
\end{equation}
A redefinition of the angle as $\hat{\phi} = \frac{r_\star (4\hat{\nu} + 6)}{2L^2 \sqrt{1+\hat{\nu}}}\;{\phi}$, where $\hat{\phi} \in [0,2\pi)$, brings the metric in polar coordinate form,
\begin{equation}
    \mathrm{d}s^2_{r,\phi} \approx \sqrt{1+\hat{\nu}\cos^2\theta} \left[ \mathrm{d}\rho^2 + \rho^2 \;\mathrm{d}\hat{\phi}^2 \right].
\end{equation}
The range in the angular coordinate that avoids conical singularities is 
\begin{equation}
{\phi} \in \left[0, \frac{4 \pi L^2 \sqrt{1+\hat{\nu}}}{ r_\star (4\hat{\nu}+6)}{}\right).\label{rangephi}
\end{equation}
In summary, the background is regular for $1\leq \xi<\infty$ with $\hat{\nu}>-1$ if we also choose the period for the ${\phi}$-coordinate to be $L_{{\phi}}= \frac{4 \pi L^2 \sqrt{1+\hat{\nu}}}{ r_\star (4\hat{\nu}+6)}{}$.
\\
These conditions for regularity encountered in backgrounds of class I do repeat in the backgrounds of classes  II and III discussed below.


\subsection{Backgrounds of class  II}
%
%
%
%
In this section we study a family of solutions to the equations of motion of eleven dimensional supergravity. We use the eleven coordinates $[t,z,w,\phi,r, \theta,\varphi,\chi,y,v_1,v_2]$. The background is written in terms of functions of the coordinates $(r,y,v_1,v_2)$. The functions are $\hat{\lambda}(y,v_1,v_2) $, $D_0(y,v_1,v_2)$, $F(r)$,     $X(r)=\lambda^2(r)$, with $F(r)$ and $ \lambda(r)$ defined in eq.(\ref{lionel1}). The functions $A_{1\phi}(r), A_{3\phi}(r)$  also appear in the background as defined in eq.(\ref{gauge_fields}).

To compactly write this family of backgrounds, we first define  the following quantities,
\begin{eqnarray}
& & ds_5^2=\frac{r^2\lambda(r)^2}{L^2} \left(-dt^2+ dz^2+ dw^2+ L^2 F(r) d\phi^2 \right) +\frac{dr^2}{r^2\lambda(r)^{{4}} F(r)},
\nonumber\\[5pt]
& & \mu_1= \sin\theta \cos\varphi,~~~~~~~\mu_2=\sin\theta \sin\varphi,~~~~~~~\mu_3= \cos\theta,
\nonumber\\[5pt]
& &D\mu_i= \left(d\mu_1+ 2\mu_2 A_{1\phi} d\phi\right)\delta_{i,1}+\left(d\mu_2 -2\mu_1 A_{1\phi}d\phi \right)\delta_{i,2}+ d\mu_3 \delta_{i,3},
\nonumber\\[5pt]
& & 2a_1= \partial_{v_2}D_0 dv_1-\partial_{v_1}D_0 dv_2,~~~~~~D\chi=d\chi+ a_1 + A_{3\phi}d\phi.\label{betoalonso}
\end{eqnarray}
After this long preamble, the eleven dimensional metric reads,
\begin{eqnarray}\label{ds11_LLM}
& &        \frac{ds^2_{11}}{\kappa^{2/3}} = e^{2\hat\lambda} \left[1+y^2e^{-6\hat\lambda}(X^3-1)\right]^{1/3}
\Big\{
\frac{4}{X}ds^2_5 + \frac{y^2e^{-6\hat\lambda}}{\left[1+y^2e^{-6\hat\lambda}(X^3-1)\right]} \mathrm{D}\mu^i\mathrm{D}\mu^i +\nonumber\\
& & \frac{4X^3}{\left[1+y^2e^{-6\hat\lambda}(X^3-1)\right]}\frac{\mathrm{D}\chi^2}{(1-y\partial_y D_0)} -\frac{\partial_yD_0}{y}dy^2-\frac{\partial_ye^{D_0}}{y}(dv_1^2+dv_2^2)\Big\}.
\end{eqnarray}
The number $\kappa$ is related to the eleven dimensional Newton constant. The  functions $\hat{\lambda}(y,v_1,v_2)$ and $D_0(y,v_1,v_2)$ are related by 
\begin{equation}
    e^{-6\hat\lambda} = -\frac{\partial_y D_0}{y(1-y\partial_y D_0)},\nonumber
\end{equation}
and $D_0(y,v_1,v_2)$ (which determines the family of solutions), satisfies the three dimensional Toda equation 
\begin{equation}
(\partial_{v_1}^2 + \partial_{v_2}^2)D_0 + \partial_y^2 e^{D_0} = 0.\label{todaeq}
\end{equation}

To complete the background, we should give the  four-form $G_4$. This is compactly written in terms of three two-forms $G = dB= d\left[ A_{3\phi}(r) d\phi\right]$, $ \mathcal{F}^{(3)} ={\sqrt{2}} d\left[A_{1\phi}(r) d\phi\right]$, and  the volume element $\mathrm{vol}\tilde{\mathbb{S}}^2 = \frac{1}{2}\epsilon_{ijk}\mu^i\mathrm{D}\mu^j\mathrm{D}\mu^k$. Using these quantities we write
\begin{equation}
    G_4 = \tilde{G}_4 + G\wedge \beta_2+\mathcal{F}^{(3)}\wedge \beta_2^{(3)}  + \star_5 \mathcal{F}^{(3)}\wedge\beta_1^{(3)}.
\end{equation}
Where we defined
\begin{equation}
    \begin{split}
        \tilde{G}_4 = -\frac{\kappa}{4}\mathrm{vol}
        \tilde{\mathbb{S}}^2\wedge&\left\{8 d\left[ \frac{y(1-y^2e^{-6\hat{\lambda}})}{1+y^2e^{-6\hat{\lambda}}(X^3-1)}-y\right]\wedge \mathrm{D}\chi\right. \\
        &\left.+ 8\frac{y(1-y^2e^{-6\hat{\lambda}})}{1+y^2e^{-6\hat{\lambda}}(X^3-1)}d a_1 -4\partial_y e^{D_0} dv_1\wedge dv_2\right\},
    \end{split}
\end{equation}
\begin{equation}
    \beta_2 = {2\kappa} \frac{X^3y^3e^{-6\hat{\lambda}}}{1+y^2e^{-6\hat{\lambda}}(X^3-1)}\mathrm{vol}\tilde{\mathbb{S}}^2,
\end{equation}
\begin{equation}
    \begin{split}
        \beta_2 ^{(3)} = \sqrt{8}\kappa&\left\{ \left[\mu^3dy + \frac{y(1-y^2e^{-6\hat{\lambda}})}{1+y^2e^{-6\hat{\lambda}}(X^3-1)}\mathrm{D}\mu^3\right]\wedge \mathrm{D}\chi\right.\\
        &\left.+ \frac{1}{2}\mu^3 \partial_y  e^{D_0} dv_1\wedge dv_2\right\},
    \end{split}
\end{equation}
\begin{equation}
    \beta_1^{(3)} = -\frac{\kappa\sqrt{8}}{X^2}d(y\mu^3).\label{finalLLM}
\end{equation}
We have checked that this background satisfies the equations of motion of eleven dimensional supergravity. In analogy with the background  of class I of Section \ref{background2b}, for $q_1=q_2=Q$ four supercharges are preserved. The structure of singularities of the backgrounds of class II is the same as that discussed in Section \ref{singularity_section}.
In fact, notice that the functions defining the solution $F(r),\lambda(r), A_{1\phi}, A_{3\phi}$ are those in eqs.(\ref{lionel1})-(\ref{gauge_fields}).

One can analyse the charge of M5-brane in this family of backgrounds. There are in fact two four-cycles on which we can integrate $G_4$ to obtain quantised fluxes. These are a four cycle formed by the interval $0\leq y\leq N \times S^2\times S^1_\chi$. The integral of $G_4$ on that cycle counts the number of colour $M5$-branes, that extend on Mink$_4$ and wrap the compact Riemann surface parametrised by $[v_1,v_2]$. The other cycle is formed by $S^2$ and a two manifold formed out of the coordinates $(v_1,v_2,y)$. The flux of $G_4$ on this four manifold counts the number of flavour (non-compact) M5 branes that extend on Mink$_4\times R_r\times S^1_\chi$.

The reader familiar with Gaiotto-Maldacena backgrounds \cite{Gaiotto:2009gz} should recognise that the family of backgrounds of class II represents the dual to a certain  deformation of a generic Gaiotto CFT \cite{Gaiotto:2009we}. We discuss the deformation of the CFT in Section \ref{sectionQFT}. 
A few words about the dual Gaiotto CFTs might help some readers. The CFT description is in terms of (at least) two stacks of M5 branes. Colour M5's that extend on the Minkowski directions (represented by $[t,z,w,\phi]$) and the directions of a compact Riemann surface, represented by a hyperbolic plane $(v_1,v_2)$. Other stacks of M5 branes extend over Minkowski$_4$ and touch the Riemann surface at a point. These are 'non-compact' (extending over the radial direction of AdS$_5$) and correspond to flavour branes. These are the two types of branes whose charges we alluded to above. Gaiotto's description of the system \cite{Gaiotto:2009we} is in terms of a punctured Riemann surface. These CFTs are typically non-lagrangian. Using the family of holographic backgrounds of Gaiotto and Maldacena is the best way to compute observables (some algebraic methods are available too). We are performing a deformation on such CFTs. This triggers a flow whose characteristic physical quantities can only be studied using the holographic dual in eqs.(\ref{ds11_LLM})-(\ref{finalLLM}).

Let us now present the third class of backgrounds. This is a family of Type IIA solutions obtained after a reduction of the backgrounds in class II. In fact, when there is rotational symmetry in the $(v_1,v_2)$-plane we can write
$v_1=\rho \cos\beta,~v_2=-\rho\sin\beta$, one can 'change basis' from a system described by the coordinates  $[r, y,\rho,\beta]$ and the function $D_0(y,\rho)$ solving a Toda equation (\ref{todaeq}), into another system described by the coordinates $[r,\sigma, \eta]$ and a function $V(\sigma,\eta)$ solving a Laplace-like equation. The change is described in \cite{Gaiotto:2009gz} and reads
\begin{equation}
 \sigma^2= \rho^2 e^{D_0(y,\rho)}, ~~y=\dot{V}=\sigma\partial_\sigma V(\sigma,\eta), ~~\log\rho= V'=\partial_\eta V(\sigma,\eta).   
\end{equation}
The Toda equation (\ref{todaeq}) is replaced by
\begin{equation}
 \sigma\partial_\sigma \left(\sigma\partial_\sigma V \right) +\sigma^2\partial^2_\eta V= \ddot{V}(\sigma,\eta)+ \sigma^2V''(\sigma,\eta)=0.\label{laplaceeq}
\end{equation}
This Laplace-like equation should be complemented with boundary and initial conditions. We briefly discuss these below. Now, we present the full Type IIA family of backgrounds, obtained after a reduction on the isometric $\beta$-direction.


\subsection{Background of class III}

The family of backgrounds in class III is defined in terms of the ten coordinates ($t$,$w$,$z$, $\phi$, $r$, $\theta$, $\varphi$, $\chi$, $\sigma$, $\eta$). The warp factors in the metric and fluxes depend on the coordinates $[r,\sigma,\eta]$. As anticipated above, the function $V(\sigma,\eta)$, and its derivatives $\dot{V}=\sigma\partial_\sigma V(\sigma,\eta)$ and $V'=\partial_\eta V(\sigma,\eta)$, solving a Laplace-like equation (\ref{laplaceeq}) play an important role. Also, the functions $F(r)$, $X(r)=\lambda^2(r)$, $A_{1\phi}(r), A_{3\phi}(r)$ defined in eqs.(\ref{lionel1})-(\ref{gauge_fields}) appear in the definition of this third  family of backgrounds.

To compactly write the family of class III backgrounds we define the functions
\begin{equation}\label{IIA_functions}
    \begin{split}
        & \tilde{f}_1=\kappa^{2/3}\left(\frac{\dot{V}\tilde{\Delta}}{2V^{\prime \prime}}\right)^{1/3},\quad \tilde{f} = X^{-1} Z=\frac{Z}{\lambda^2(r)},\quad \tilde{f}_2=\frac{2\dot{V}V^{\prime\prime}}{Z^2\tilde{\Delta}}, \quad \tilde{f}_3 = \frac{4X^3\sigma^2V^{\prime\prime}}{2X^3\dot{V}-\ddot{V}}Z \, ,
        \\
        &\tilde{f}_4=\frac{2V^{\prime\prime}}{\dot{V}}Z,\quad \tilde{f}_5 = \frac{2(2X^3\dot{V}-\ddot{V})}{Z^2\dot{V}\tilde{\Delta}}, \quad  \tilde{f}_6 = \frac{2X^3\dot{V}\dot{V}^{\prime}}{2X^3\dot{V}-\ddot{V}} ,
        \\
        &\tilde{f}_7=\frac{2X^3\dot{V}^2V^{\prime\prime}}{\tilde{\Delta}+2V^{\prime\prime}\dot{V}(X^3-1)},\quad \tilde{f}_8=\frac{\dot{V}\left[(\dot{V}^{\prime})^2-V^{\prime\prime}\ddot{V}\right]}{\tilde{\Delta}+2V^{\prime\prime}\dot{V}(X^3-1)},\quad \tilde{f}_9=\frac{\sigma V^{\prime\prime}}{(\dot{V}^{\prime})^2-V^{\prime\prime}\ddot{V}}, \\
        &
        \tilde{\Delta}=(\dot{V}^{\prime})^2+V^{\prime\prime}(2\dot{V}-\ddot{V}) ,\quad Z=\left[ \frac{(\dot{V}^{\prime})^2+V^{\prime\prime}(2X^3\dot{V}-\ddot{V})}{(\dot{V}^{\prime})^2+V^{\prime\prime}(2\dot{V}-\ddot{V})}\right]^{1/3}, \quad g(\sigma,\eta) = \frac{\dot{V}^\prime}{\ddot{V}V^{\prime\prime} - (\dot{V}^\prime)^2}.
    \end{split}
\end{equation}
We use the definitions of 
$D\mu_i$ and the five-dimensional metric $ds_5^2$ in eq.(\ref{betoalonso}), together with the one-form  $B=A_{3\phi} d\phi$ and the two-forms $ \mathcal{F}^{(3)} = {\sqrt{2}}\;\mathrm{d}\left[A_{1\phi}(r) \mathrm{d}\phi\right]$ and 
$G = dB$.
The Type IIA backgrounds of class III read in string frame,

\begin{equation}
    \begin{split}
        &\mathrm{d}s^2_{10}= \tilde{f}_1^{\frac{3}{2}} \tilde{f}_5^{\frac{1}{2}} \left[ 4\tilde{f} \mathrm{d}s^2_5 + \tilde{f}_2 \mathrm{D}\mu_i\mathrm{D}\mu^i + \tilde{f}_3(\mathrm{d}\chi+B)^2 + \tilde{f}_4(\mathrm{d}\sigma^2 + \mathrm{d}\eta^2) \right]\\
        &e^{\frac{4}{3}\Phi} = \tilde{f}_1\tilde{f}_5, \quad C_1=\tilde{f}_6(\mathrm{d}\chi + B),\\
    &H_3=\mathrm{d}B_2=2\mathrm{vol}\tilde{\mathbb{S}}^2\wedge\Big[-g\,\mathrm{d}\tilde{f}_7+\tilde{f}_8\mathrm{d}g+\tilde{f}_9(\sigma V^{\prime\prime}\mathrm{d}\eta+\dot{V}^{\prime}\mathrm{d}\sigma) \Big]\\
    &+\sqrt{8}\mathcal{F}^{(3)}\wedge \left[ \mu^3\tilde{f}_9(\sigma V^{\prime\prime}\mathrm{d}\eta+\dot{V}^{\prime}\mathrm{d}\sigma)-g\left(\tilde{f}_8\mathrm{d}\mu^3 + \mu^3\mathrm{d}\dot{V} \right)   \right],\\
        &F_4=\mathrm{d}C_3-H_3\wedge C_1=-2\tilde{f}_7\mathrm{vol}\tilde{\mathbb{S}}^2\wedge G -\frac{\sqrt{8}}{X^2}\star_{5}\mathcal{F}^{(3)}\wedge\mathrm{d}(\mu^3\dot{V})\\
        &+2 \mathrm{vol}\tilde{\mathbb{S}}^2\wedge \Big[\left(1+g \,\tilde{f}_6\right)\mathrm{d}\tilde{f}_7-\tilde{f}_6\tilde{f}_8\mathrm{d}g-\tilde{f}_6\tilde{f}_9(\sigma V^{\prime\prime}\mathrm{d}\eta+\dot{V}^{\prime}\mathrm{d}\sigma)  \Big]\wedge (\mathrm{d}\chi+B)\\
        &+\sqrt{8}\mathcal{F}^{(3)}\wedge\Big[ - \mu^3\tilde{f}_6\tilde{f}_9(\sigma V^{\prime\prime}\mathrm{d}\eta+\dot{V}^{\prime}\mathrm{d}\sigma)+ (1+g\tilde{f}_6)\left(\tilde{f}_8\mathrm{d}\mu^3 + \mu^3\mathrm{d}\dot{V} \right)\Big]\wedge(\mathrm{d}\chi+B).\label{10dGM-2}
    \end{split}
\end{equation}

We have checked that the background above solves the equations of motion of Type IIA.

As discussed in detail in \cite{Gaiotto:2009gz,Macpherson:2024frt,Lozano:2016kum,Aharony:2012tz,Reid-Edwards:2010vpm, Nunez:2019gbg}, the function $V(\sigma,\eta)$ is determined by the dual UV CFT. For this family of backgrounds in eq.(\ref{10dGM-2}), we are studying the 
dual description of a linear quiver ${\cal N}=2$ four dimensional CFT deformed by a VEV. 

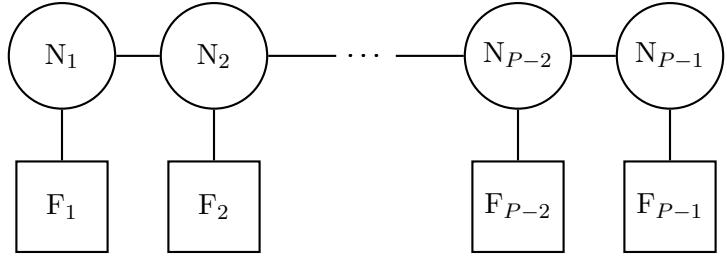
\begin{figure}[ht]
\begin{center}
	\begin{tikzpicture}
	\node (1) at (-4,0) [circle,draw,thick,minimum size=1.4cm] {N$_1$};
	\node (2) at (-2,0) [circle,draw,thick,minimum size=1.4cm] {N$_2$};
	\node (3) at (0,0)  {$\dots$};
	\node (5) at (4,0) [circle,draw,thick,minimum size=1.4cm] {N$_{P-1}$};
	\node (4) at (2,0) [circle,draw,thick,minimum size=1.4cm] {N$_{P-2}$};
	\draw[thick] (1) -- (2) -- (3) -- (4) -- (5);
	\node (1b) at (-4,-2) [rectangle,draw,thick,minimum size=1.2cm] {F$_1$};
	\node (2b) at (-2,-2) [rectangle,draw,thick,minimum size=1.2cm] {F$_2$};
	\node (3b) at (0,0)  {$\dots$};
	\node (5b) at (4,-2) [rectangle,draw,thick,minimum size=1.2cm] {F$_{P-1}$};
	\node (4b) at (2,-2) [rectangle,draw,thick,minimum size=1.2cm] {F$_{P-2}$};
	\draw[thick] (1) -- (1b);
	\draw[thick] (2) -- (2b);
	\draw[thick] (4) -- (4b);
	\draw[thick] (5) -- (5b);
	\end{tikzpicture}
\end{center}
\caption{Long quiver of length ($P-1$) with gauge nodes $N_i$ and flavour nodes $F_i$. The quiver is \textit{balanced and conformal} if $F_i = 2 N_i - N_{i-1}-N_{i+1}$.}
\label{fig:quiverfiga}
\end{figure}

Let us briefly summarise how the function $V(\sigma,\eta)$ is determined in terms of the data of the linear quiver field theory. Consider a quiver of the form displayed in Figure \ref{fig:quiverfiga}.
The information about the number of colours (gauge groups) and flavours (global groups) is encoded in the so-called rank function,
\begin{equation}
 {\cal R}(\eta) = \begin{cases} 
       N_1 \eta & \ , \quad  0\leq \eta \leq 1 \\
         N_l+ (N_{l+1} - N_l)(\eta-l) & \ , \quad  l \leq \eta\leq l+1\ ,\;\;\; l:=1,\ldots, P-2\\\label{rankfull}
 %
         N_{P-1}(P-\eta) & \ , \quad  (P-1)\leq \eta\leq P 
      \end{cases} \ .
\end{equation}
In fact, the rank of the $j^{th}$ gauge group is the value of the rank function at $\eta=j$. The number of flavours $F_i$ is encoded in the second derivative  ${\cal R}''(\eta)$,
\begin{equation}
{\cal R}''(\eta)=\sum_{j=1}^{P-1} (2N_j -N_{j-1} -N_{j+1})\delta(\eta-j)= \sum_{j=1}^{P-1} F_j\delta(\eta-j) \ .\label{rsecond}
\end{equation}
The field theory data (the quiver) is encoded in the supergravity background via the initial condition of the Laplace equation (\ref{laplaceeq}). 
In fact, the complete system to  solve is
\begin{eqnarray}
& & \sigma\partial_\sigma\left( \sigma\partial_\sigma V \right)+ \sigma^2\partial^2_\eta V= 0,\label{laplacesystem}\\
& & V(\sigma\to\infty,\eta)=0, ~~V(\sigma,\eta=0)=V(\sigma,\eta=P)=0,~~\sigma\partial_\sigma V(\sigma=0,\eta)={\cal R}(\eta).\nonumber
\end{eqnarray}
The solution to this partial differential equation is,
\begin{equation}
V(\sigma,\eta)=-\sum_{k=1}^\infty {\cal R}_k 
\sin \left[\frac{k \pi\eta}{P}\right] 
K_0\left[ \frac{k \pi \sigma}{P}\right] \quad {\rm with} 
\quad {\cal R}_k=\frac{2}{P}\int_0^P {\cal R}(\eta) 
\sin \left[ \frac{k \pi\eta}{P}\right] d\eta.\label{Vsigmaeta}   
\end{equation}
To close the presentation of the third class of backgrounds, we emphasize that the structure of singularities is the same as that in backgrounds I and II, namely the background is smooth as long as the parameter $\hat{\nu}>-1$, the period of the angle $\phi$ is taken as in eq.(\ref{rangephi}), with the coordinate $r$ ranging in $[r_\star,\infty)$ (equivalently $1\leq \xi<\infty)$. The interesting characteristic of class III backgrounds is that there are also localised singularities in the $\eta$-coordinate (at $\sigma=0$). These appear at each point in which the rank function changes slope, equivalently, at the position in which the flavour D6 branes are localised. In fact, careful study indicates that stacks of  $F_j=2N_j-N_{j+1}-N_{j-1}$ D6 branes sit at the values of $\eta$ where slope changes in the rank function, see eq.(\ref{rsecond}).
In fact, integrating the Page fluxes of $\hat{F}_4= F_4-B_2\wedge F_2$ over the cycle $k\leq \eta\leq(k+1) \times S^2\times S^1_\chi$ we find the number of D4 branes in each interval of $\eta$. Integrating $F_2$ on $[\eta,S^1_\chi]$ counts the number of D6 branes due to changes in the slope of the rank function. Finally, integrating $H_3$ over the three cycle $[\eta,S^2]$ counts the number of NS-five branes.

{
\subsection{How are these backgrounds constructed?}\label{rough-view}
In this section we discuss how the backgrounds of Type I, II and III are constructed. The section is short and should be thought of as an advance of  the more complete treatment 
that we present in \cite{CHINZ2}.
\\
The way in which the Type IIB, eleven dimensional supergravity and Type IIA are constructed starts from a solution in five dimensions. 
The five dimensional gauged supergravity is found after compactifying type IIB supergravity on $S^5$  \cite{Cvetic:1999xp}. It  contains in its bosonic sector the metric, three gauge fields $A^i$, as well as two scalar fields denoted as $\vec{\Phi}=(\Phi_1,\Phi_2)$. The action reads,
\begin{eqnarray}
 & &    \mathrm{S} = \frac{1}{2\kappa}\int \mathrm{d}^5x \sqrt{-g}\Bigg[R - \frac{1}{2}(\partial\Phi_1)^2- \frac{1}{2}(\partial\Phi_2)^2  - \frac{1}{4}\sum _{i=1}^3X_i^{-2}F^i_{\mu\nu}F^{i\, \mu\nu} + \frac{1}{4}\epsilon^{\mu\nu\rho\sigma\lambda}A_{\mu}^1F_{\nu\rho}^2F_{\sigma\lambda}^3\nonumber\\
 & &~~~~~~~~~~~~~~~~~~~~~~~~~+ \frac{4}{L^2}\sum_{i=1}^3X_{i}^{-1}\Bigg],\label{5d_gauged_action}
\end{eqnarray}
with $F^i=\mathrm{d}A^i$, $X_i=e^{-\frac{1}{2}\vec{a}_i\cdot\vec{\Phi}}$ where
\begin{equation}
    \vec{a}_1 = \left(\frac{2}{\sqrt{6}},\sqrt{2}\right), \quad \vec{a}_2 = \left( \frac{2}{\sqrt{6}},-\sqrt{2}\right), \quad \vec{a}_3 = \left(-\frac{4}{\sqrt{6}},0\right).
\end{equation}
We study a consistent truncation of the above system, that admits a solution of the form,
\begin{equation}\label{5d_soliton}
    \begin{split}
        &\mathrm{d}s^2_5=\frac{r^2\lambda^2(r)}{L^2}\left( -\mathrm{d}t^2+\mathrm{d}z^2+\mathrm{d}w^2+L^2F(r)\mathrm{d}\phi^2\right)+\frac{\mathrm{d}r^2}{r^2\lambda^4(r)F(r)},\\
         &\Phi_1 \equiv \Phi =\sqrt{\frac{2}{3}}\mathrm{ln}\lambda^{-6}(r),\quad \Phi_2 = 0,\\
        & A^1 = A^2 = q_1\left[\lambda^6(r)-\lambda^6(r_{\star})\right]L\, \mathrm{d}\phi,\quad A^3 = q_2\left[\frac{1}{\lambda^6(r)} - \frac{1}{\lambda^6(r_{\star})}\right]L\, \mathrm{d}\phi,
    \end{split}
\end{equation}
where  $F(r)$ and $\lambda(r)$ are functions of the radial coordinate,
\begin{equation}\label{F_and_lambda_r}
   F(r) =\frac{1}{L^2}-\frac{\varepsilon\ell^2 L^2}{r^4}\left(q_1^2 -\frac{q_2^2}{\lambda^6(r)}\right),\quad \lambda^6(r)=\frac{r^2+\varepsilon\ell^2}{r^2}.
\end{equation}
Here $r_{\star}$ is defined to be the largest root of $F(r)$, satisfying $F(r_{\star})=0$, which expresses the end of the geometry, while $\ell$ is a parameter and $\varepsilon$ is just a sign indicating two non-diffeomorphic branches of the supergravity solution \cite{Anabalon:2024che}. This configuration of fields brings \eqref{5d_gauged_action} to the form:
\begin{equation}
    \mathrm{S} = \frac{1}{2\kappa}\int d^5x \sqrt{-g}\left[R - \frac{1}{2}(\partial\Phi)^2+ \frac{4}{L^2}\left(\frac{2}{X}+X^2\right) - \frac{1}{2X^2}F_{\mu\nu} ^{1} F^{1\, \mu\nu} - \frac{1}{4}X^4F^3_{\mu\nu} F^{3\, \mu\nu}\right],
\end{equation}
where $X=e^{-\Phi/\sqrt{6}}$. We have verified that the configuration satisfies all the  equations of motion.
The next step is to embed this solution into either Type IIB or in eleven dimensional supergravity. Fortunately, these lifts have been constructed in  \cite{Cvetic:1999xp}  and in \cite{Gauntlett:2007sm}. These lifts display an interesting difference: the lift to IIB \cite{Cvetic:1999xp} gives a unique solution, which we referred to above as the class I background. On the other hand, the lift in \cite{Gauntlett:2007sm} provides an infinite family of solutions, one for each possible solution of the function $D(y, v_1,v_2)$--we called this family class II backgrounds. In the case that we have rotational symmetry in the $(v_1,v_2)$-plane, we reduce to Type IIA, generating the infinite family of backgrounds in class III.  As a generic comment, we note that the role played by the five dimensional scalar field $\Phi$ is to change the warp factors of the lifted backgrounds. The role of the gauge fields is to twist (or fiber) the five-dimensional manifold with the internal space, associated with the R-symmetry of the dual field theory. }
\\
Let us now discuss some field theoretical aspects of these backgrounds.


\section{Field Theory}\label{sectionQFT}

\subsection{Some aspects of the dual QFTs}\label{aspectsQFT}

As we discussed around eqs.(\ref{zizu})-(\ref{zivu}) we can think about our backgrounds as deformation of those in \cite{Anabalon:2021tua,Chatzis:2024top,Chatzis:2024kdu}. In the very far UV ($\xi\to\infty$) and in the scaling of parameters in eq.(\ref{zivu}), choosing $\hat{q}\to 0$, we have AdS$_5\times S^5$ or $\cN=4$ SYM for the backgrounds in class I. First we discuss the dual QFT to backgrounds in class I, then we mention classes II and III. 

The CFT $\cN=4$ SYM is deformed by the VEV of a dimension three operator that preserves SUSY (the operator is identified in \cite{Castellani:2024ial}). This is accompanied by a twisted-compactification along the direction $\phi$. Holographically, the 'twisted' character of the compactification is realised by the one-forms $A_1, A_2,A_3$ in eq.(\ref{zivu}), that mix  a part of the Lorentz group (an SO(2) represented by $\phi$-translations) with the R-symmetry; a $U(1)^3$ part of SO(6)$_R$ represented holographically by the directions $\phi_1,\phi_2,\phi_3$.

On top of this we switch on a VEV for an operator of dimension two. Switching on this operator in the QFT is represented in gravity by the parameter $ \ell$ and the new functions $F(\xi),\lambda(\xi)$ and the different functional dependence of the one forms $A_1,A_2,A_3$. In the paper \cite{CHINZ2}, we provide the required boundary expansions to justify this. 

The deformation by the operator of dimension two is the Coulomb branch deformation of \cite{Freedman:1999gk}. These flows are typically accompanied by a singularity deep in the bulk which can be understood by a continuous distribution of branes.
In our case, as in the case of \cite{Anabalon:2024che} we avoid this singularity by closing the space (analogously gapping the QFT) before reaching it.

The twisted compactification together with the Coulomb branch deformation generates dynamics that is holographically represented by backgrounds in class I. 
\\
For backgrounds in classes II and III the logic is similar.
In this case, in the far UV we have $N=2$ SCFTs in four dimensions, as the backgrounds asymptote to AdS$_5\times M$ (here $M$ is the  internal six space in class II and internal five space in class III). For class II, the dual CFTs are non-Lagrangian Gaiotto CFTs \cite{Gaiotto:2009we} and for class III are $N=2$ linear quiver CFTs in four dimensions.

Both in the case of classes II and III backgrounds, the dual field theories admit the same deformations as discussed above, in particular the twisted compactification reducing the system to $(2+1)$-dimensions  relies only on the existence of a preserved R-symmetry \cite{Cassani:2021fyv, Kumar:2024pcz}.
Note that in this work we switch-on another deformation that makes the QFT explore the Coulomb branch.

{Let us be slightly more precise about the operators obtaining a VEV. We can be more explicit in the case the UV-CFT is ${\cal N}=4$ SYM. In this case we have a perturbative (Lagrangian) description of the system. We cannot be more explicit in the case of the field theories dual to class II backgrounds, as the theory is generically non-Lagrangian. For class III backgrounds, we expect things to behave similarly to the case of class I, repeated for each node of the quiver, with the additional subtlety that the bi-fundamental matter plays a role in the operators. 
\\
A direct application of holographic renormalisation indicates that a dimension two operator and a dimension three operator acquire  VEVs. The dimension three operator is discussed in the paper \cite{Castellani:2024ial}. In that case is a current operator that for this  deformation of Super-Yang-Mills is ${\cal O}_3=\text{Tr} \lambda^a \gamma_\mu \lambda^a$. This operator is  descendant from a chiral operator. One can choose among the four fermions in ${\cal N}=4$ SYM a combination that breaks from $SU(4)\to U(1)^3$. The $U(1)^3$ is reflected in the three isometries $\partial_{\phi_i}$ of the class I  background. The reduction in R-symmetry comes together with a  reduction of the amount of SUSY preserved. In fact, our backgrounds preserve only four supercharges.
\\
Let us comment about the operator of dimension two acquiring a VEV. This operator was studied in  \cite{Freedman:1999gk}. There, it was shown that the operator is  the symmetrised product ${\cal O}_2=\text{Tr}X_{(i } X_{j ) }$. This operator transforms in the $(0,2,0)$ representation of $SU(4)_R$ and it is associated with deformations of the metric and of the five form. The operator getting a VEV moves us along the Coulomb branch of the CFT, generating (in the un-gapped case, see below) a singularity in the metric that is associated with a continuous distribution of D3 branes. 
}
\\
The interest of the families of backgrounds in classes I, II, III is that the characteristic singularity of the Coulomb branch solutions does not show-up, by ending the space (the range of the radial direction $1\leq \xi<\infty$) before reaching the singularity. This is analogous to introducing a gap in the QFT. The parameter $\hat{\nu}$ controls how far away from the end of the space on the $\xi<1$ side, the singular behaviour 'is located'. For $\hat{\nu}\to -1^+$  (and $\theta\sim 0$ in class I) the effects of the singularity can make themselves apparent by the need to correct the background due to the large values of the 
invariants like $R_{ab}^2, R_{abcd}^2$. Field theoretically we interpret this by thinking that the gap in the QFT is not far enough from the mass of the light fields we would need to introduce, had we not avoided the singularity.

Let us briefly elaborate a bit more on this. The Coulomb branch solution of \cite{Freedman:1999gp}
 is singular. Hence, the description of the QFT-dynamics provided by supergravity is incomplete, and the singularity needs to be resolved. On the QFT side, this suggests that some light fields (appearing at the Coulomb VEV scale) are not considered by the holographic description, which then breaks down.

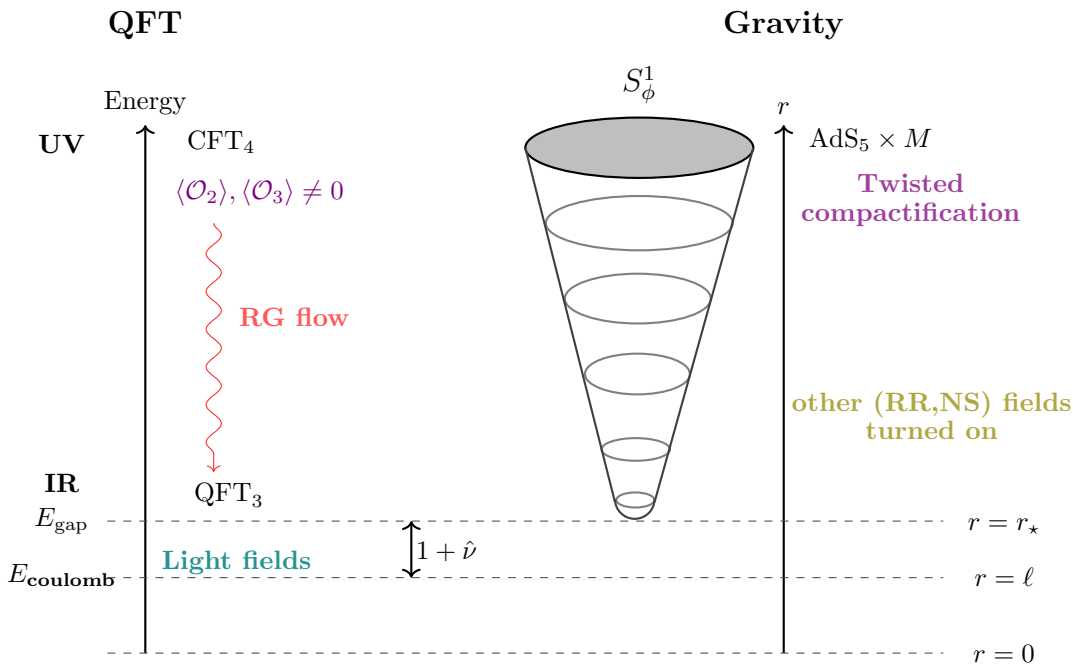
\begin{figure}[h!]
\begin{tikzpicture}

\def\a{1.5}        
\def\b{0.4}        
\def\h{5}          
\def\r{0.3}        

\pgfmathsetmacro{\hs}{\h - \r}

\draw[thick,fill=lightgray]
  (0,0) ellipse ({\a} and {\b});
  \draw[thick,black!50] (0,-1) ellipse ({\a-0.27} and {\b-0.04});
\draw[thick,black!50] (-0.02,-2) ellipse ({\a-0.54} and {\b-0.07});
\draw[thick,black!50] (-0.025,-3) ellipse ({\a-0.81} and {\b-0.13});
\draw[thick,black!50] (-0.047,-4) ellipse ({\a-1.05} and {\b-0.25});
\draw[thick,black!50] (-0.052,-4.67) ellipse ({\a-1.25} and {\b-0.30});

\draw[thick,black!75] (-\a, 0) -- (-\r-0.02, -\hs);
\draw[thick,black!75] (\a, 0) -- (\r-0.1, -\hs);

\begin{scope}[rotate around={180:(-\r+0.25,-\hs+0.02)}]
\draw[thick,,black!75]
  (-\r, -\hs) arc[start angle=180, end angle=0, radius=\r-0.04];
\end{scope}

\draw[->, thick] (\a + 0.4, -\hs-2) -- (\a + 0.4, -\hs + 5)
  node[above] {\small$r$};

\node[right] at (\a + 0.6, -\hs + 4.8) {\small$\mathrm{AdS}_5 \times M$};

\node[above] at (0, \b + 0.05) {\large$S^1_\phi$};

\draw[->,red!85, decorate,decoration={snake,amplitude=1mm,segment length=5mm}](\a - 7.1, -\hs + 3.7)-- (\a - 7.1, -\hs+0.4);
\node[above right] at (\a-6.9,-\hs+2.2) {\bf\color{red!65}\small RG flow};

\node[above] at  (-\a - 3.9,  -\hs-0.2) {\small $\mathrm{QFT}_3$};

\draw[->, thick] (-\a - 5, -\hs-2) -- (-\a - 5, -\hs + 5)
  node[above] {\small Energy};

\node[above] at (-\a - 5, -\hs + 6) { \large \textbf{QFT}};

\node[above] at (\a + 0.4, -\hs + 6) { \large\textbf{Gravity}};

\draw[dashed,black!70] (-\a - 5.5, -\hs-0.25 ) -- (\a + 2.5, -\hs-0.25 );
\draw[dashed,black!70] (-\a - 5.5, -\hs - 1) -- (\a + 2.5, -\hs - 1);
\draw[dashed,black!70] (-\a - 5.5, -\hs - 2) -- (\a + 2.5, -\hs - 2);

\node[above] at (\a + 3.3 , -\hs -0.55) {\small \textbf{$r=r_\star$}};
\node[above] at (\a + 3.3 , -\hs -1.25) {\small \textbf{$r=\ell$}};
\node[above] at (\a + 3.3 , -\hs -2.25) {\small \textbf{$r=0$}};

\node[above] at (-\a -6.1 , -\hs ) {\small \textbf{IR}};
\node[above] at (-\a -6.1 , -\hs+4.5 ) {\small \textbf{UV}};

\node[above] at (-\a -6.1 , -\hs -1.25) {\small \textbf{$E_{\text{coulomb}}$}};
\node[above] at (-\a -6.1 , -\hs -0.55) {\small \text{$E_{\text{gap}}$}};

\draw[<->,thick] (-\a - 1.5, -\hs - 1) -- (-\a - 1.5, -\hs - 0.25);

\node[left] at (-\a - 0.5, -\hs - 0.65) {\small $1+ \hat{\nu}$};

\node[above] at (-\a -4.0 , -\hs+4.5 ) {\small \textbf{$\mathrm{CFT}_4$}};

 \node[below] at (-\a - 3.8, -\hs - 0.5) {\bf \color{teal!85}\small Light fields};

\node
  at (\a + 2.28, -\hs + 1.1)
  {
   {\small
      \color{olive!75}
      \bf
      \def\arraystretch{.9}
      \def\tabcolsep{-5pt}
      \begin{tabular}{c}
       other (RR,NS) fields
        \\
        turned on
      \end{tabular}
    }
  };

\node[above] at (-\a - 3.5 , \hs -5.6 ) {\small \color{violet} $\langle \mathcal{O}_2\rangle, \langle \mathcal{O}_3 \rangle \neq 0$};

\node[right] at (\a + 0.4, -\hs + 4) 
 {
   {\small
      \color{violet!75}
      \bf
      \def\arraystretch{.9}
      \def\tabcolsep{-5pt}
      \begin{tabular}{c}
       Twisted 
        \\
        compactification
      \end{tabular}
    }
  };


 
\end{tikzpicture}
\caption{An overview of our twisted compactification procedure in supergravity and its realization in the dual field theory. The distance between the two scales is parametrized by $1+\hat{\nu}$.}
\label{drawing}
\end{figure}

Suppose that we gap the QFT with a new scale. If this is close or degenerate with the Coulomb VEV scale, one could expect that the light degrees of freedom mentioned above are still important for the dynamics close to the gap scale. Conversely, in gravity, higher invariants could be large (not divergent), motivating the need to correct the solution. The parameter $\hat\nu$ controls the need for such corrections. As $\hat\nu$ drifts away from $\hat\nu\approx -1^+$, to larger values
the gap and the scale of the light fields separate from the gap scale, the higher curvature invariants become more tamed and supergravity becomes trustable. We try to convey this in Figure \ref{drawing}
\\

In what follows, we discuss a set of observables that cannot be calculated by using QFT-methods.

\subsection{Wilson loops}\label{wilsonsection}

In this section we study ''universal'' Wilson loops. By this we mean a Wilson line that does not explore any of the internal dimensions of the space time. The result, as expected is dynamically equivalent in the three families of backgrounds. Only the effective tension of the string is affected by the details of the internal space. More elaborated configurations for the probe string are studied in \cite{CHINZ2}. We study the effect of the parameter $\hat{\nu}$ on the behaviour of the Wilson line. In fact, as we discussed for $\hat{\nu}\approx -1^+$ and $\theta\approx 0$ the background should be corrected by higher curvature terms (at least close to the end of the space $\xi=1$). The Wilson loop for $\hat\nu\to-1^+$ presents a  qualitatively different behaviour to that calculated with other values of $\hat\nu$ further away from the edge of the allowed window. As expected, confining behaviour is observed for values of the $\hat{\nu}$-parameter for which the background is trustable. 

When we compute the Wilson loop in the backgrounds III, there is an extra subtlety to be considered. We need to specify a value of the coordinate $\eta=\eta_\star$, that is equivalent to choosing a particular gauge group for the calculation. In this paper we do not allow the F1 string to explore the $\eta$-direction, but experience gathered in a similar system \cite{Giliberti:2024eii} suggests that the string should explore the $\eta$-direction, moving towards the closer flavour group. This is the holographic description of a system presenting screening. We do not explore this here.
\\
{\underline{A word of caution:}
In what follows we discuss {\it non}-SUSY Wilson loops. These could be perturbatively unstable. In which case, the discussion below would be devoid of content. Nevertheless, we believe that this is not the case. Let us point out main differences with similar studies in the bibliography. For example, in \cite{Avramis:2006nv} an analysis of the stability is carefully performed (fluctuations of the string along different directions of the space). In that case, the analysis shows the existence of unstable configurations. The main difference of the study in \cite{Avramis:2006nv} with ours is that the
background used in \cite{Avramis:2006nv} is the {\it singular} Coulomb branch solution. In our case we are avoiding the singularity by gapping the QFT. Fluctuations of the string soliton have better changes of being non-tachyonic.
\\
 As we discuss below, the function $L(r_0)$--the separation between the quark-anti-quark pair is multivalued only when the background needs higher order corrections (though non-singular!). The multi-valued character of $L(r_0)$  is typically characteristic of a  phase transition to a more favourable configuration. 
The only way to be more sure of the validity of our analysis is to perform the fluctuations analysis, as done for example in \cite{Avramis:2006nv} or in \cite{Chatzis:2024dlt}. We postpone this analysis to our forthcoming paper \cite{CHINZ2}.
}
\\
We start discussing rectangular Wilson loops for backgrounds of class I.
%
\subsubsection{Wilson loop for the Type IIB background of class I}
We study the Wilson loop in the background \eqref{S5-1} using the standard holographic methods \cite{Sonnenschein:1999if,Nunez:2009da,Kol:2014nqa,Nunez:2023nnl}. We consider universal embeddings, namely an open $\mathrm{F}1$ probe string with fixed endpoints at $r=\infty$, whose worldsheet coordinates we parametrize as $\tau=t$ and $\sigma=w$, with $r=r(w)$. The endpoints of the string correspond to a probe quark-anti-quark pair separated at distance $L_{\mathrm{QQ}}$ in the dual field theory, while the rest of the string enters the $r$ direction in a $U$ shaped configuration, down to a turning point $r_0$. As the point $r_0$ takes different values, expressing the string profile delving deeper into the bulk, the length of the separation $L_{\mathrm{QQ}}$ also changes. The Nambu-Goto action for the probe simplifies to
\begin{equation}
	S_{\mathrm{NG}}\propto \int_{-L_{\mathrm{QQ}}/2}^{L_{\mathrm{QQ}}/2}\mathrm{d}w\sqrt{\mathcal{F}^2+\mathcal{G}^2r^{\prime 2}},
\end{equation}
where $\mathcal{F},\mathcal{G}$ are functions that depend on the details of each background and the radial coordinate is parametrized as $r=r(w)$.  \\

\begin{figure}[t]
\centering
\begin{subfigure}{0.5\linewidth}
\includegraphics[width=\linewidth]{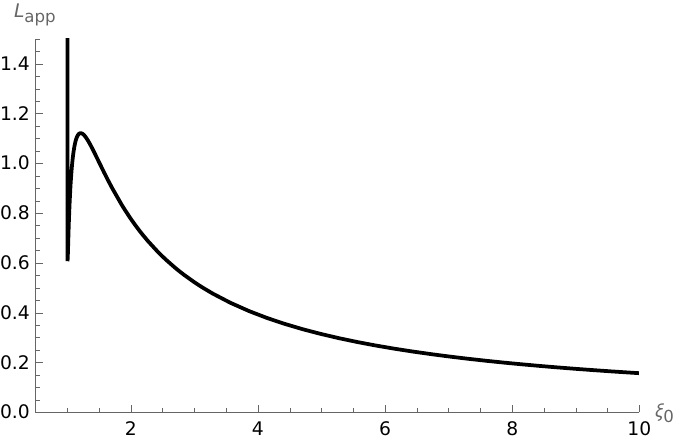}
\caption{Plot of the approximate length $L_{app}$ for  $\hat{\nu}=-0.99$.}
\label{figuraLappa}
\end{subfigure}
\hfill
\begin{subfigure}{0.44\linewidth}
\includegraphics[width=\linewidth]{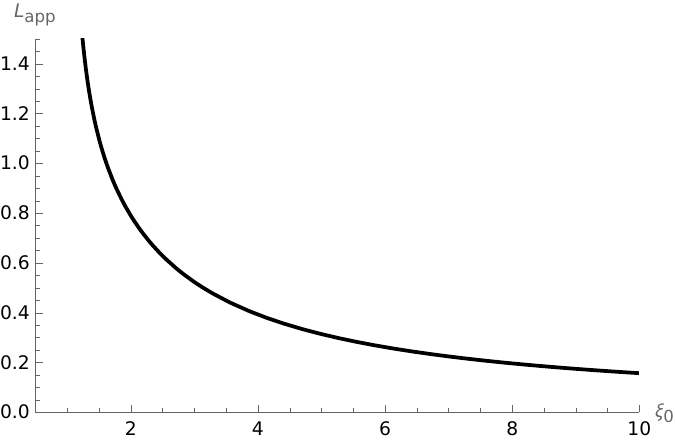}
\caption{Plot of the the approximate length $L_{app}$ for  $\hat{\nu}=1$.}\label{figuraLappb}
\end{subfigure}
\caption{Plot of the approximate separation of the quark-anti quark pair in eq.(\ref{Lapproxi}).}
\label{Figure3}
\end{figure}

Following \cite{Nunez:2009da} we have expressions for the  separation $L_{QQ}(r_0)$, together with an approximate expression L$_{app}(r_0)$ for the separation written in \cite{Kol:2014nqa}, and the energy E$_{QQ}(r_0)$ of the pair as functions of $r_0$
\begin{equation}\label{WLintegrals}
\begin{split}
	& V_{\mathrm{eff}}(r;r_0)=\frac{\mathcal{F}(r)\sqrt{\mathcal{F}^2(r)-\mathcal{F}^2(r_0)}}{\mathcal{F}(r_0)\mathcal{G}(r)},\\
	& L_{\mathrm{app}}(r_0)=\left.\frac{\pi\mathcal{G}(r)}{\mathcal{F}^{\prime}(r)}\right|_{r=r_0},\quad  L_{\mathrm{QQ}}(r_0)=2\int_{r_0}^{\infty}\frac{\mathrm{d}r}{V_{\mathrm{eff}}(r;r_0)},\\
	& E_{\mathrm{QQ}}(r_0)=\mathcal{F}(r_0)L_{\mathrm{QQ}}(r_0)+2\int_{r_0}^{\infty}\mathrm{d}r~ \frac{\mathcal{G}(r)}{\mathcal{F}(r)}\sqrt{\mathcal{F}^2(r)-\mathcal{F}^2(r_0)}-2\int_{r_{\star}}^{\infty}\mathrm{d}r~\mathcal{G}(r).
\end{split}
\end{equation}
Here $r_{\star}$ denotes the point at which the space terminates in a smooth fashion.  
As anticipated above, we work with the simplest and universal embedding for the probe string, where only the radial coordinate has a profile $r=r(w)$ and the rest of the coordinates are kept fixed. In more detail our configuration is,
\begin{equation}
 t=\tau,~~w=\sigma,~~r=r(w),~~\theta_0=0 ~~\text{or}~~\theta_0=\frac{\pi}{2}.   
\end{equation}
We have checked that these embeddings solve the full equations of motion for the string. In particular notice that there is an embedding for $\theta_0=0$, which  for values of $\hat\nu\approx -1^+$ might detect the effects of large curvature corrections. We come back to this below.
\\
The induced metric of the string is
\begin{equation}
    \mathrm{d}s^2_{\text{ind}}=-\frac{r^2\zeta(r,\theta_0)}{L^2}\mathrm{d}t^2 + \left[ \frac{r^2\zeta(r,\theta_0)}{L^2}+ \frac{\zeta(r,\theta_0)r^{\prime2}}{r^2F(r)\lambda^6(r)}\right]\mathrm{d}w^2,
\end{equation}
from which we calculate the action,
\begin{eqnarray}
& &     \mathrm{S}_{\mathrm{F}1} = \frac{1}{2\pi}\int \mathrm{d}t\mathrm{d}w\sqrt{-\mathrm{det}g_{\text{ind}}}=\frac{\mathcal{T}}{2\pi}\int_{-L_{\mathrm{QQ}}/2}^{L_{\mathrm{QQ}}/2}\mathrm{d}w\sqrt{\mathcal{F}^2+\mathcal{G}^2r^{\prime2}},\nonumber\\
 & &    \mathcal{F} = \frac{r^2\zeta(r,\theta_0)}{L^2},\quad \mathcal{G}= \frac{\zeta(r,\theta_0)}{L\lambda^3(r)\sqrt{F(r)}}. \label{F_and_G_WL_IIB_embeddingI}
\end{eqnarray}
The approximate length takes the following expression: 
\begin{equation}
    L_{\mathrm{app}}(r_0)=\frac{\pi L\zeta(r_0,\theta_0)}{r_0\lambda^3(r_0)\sqrt{F(r_0)}\left[2\zeta(r_0,\theta_0)+r_0\partial_{r_0}\zeta(r_0,\theta_0)\right]}.\label{Lapproxi}
\end{equation}
We plot $L_{app}(r_0)$ in Figures \ref{figuraLappa} and \ref{figuraLappb} for two values of the parameter $\hat\nu$ and generic values $q_1=q_2=Q=1$, $L=\ell=1$, $\theta_0={0}$. These figures display a qualitatively different behaviour
for the quantity in eq.(\ref{Lapproxi}). In fact, close to the edge of the allowed values ($\hat\nu=-0.99$) we find $L_{app}$ to be double valued. On the other hand, for a value of $\hat\nu$ further  from the edge the function $L_{app}$ is single valued. In contrast, the  plot of $L_{app}$ for $\theta_0=\frac{\pi}{2}$ and for $\hat{\nu}=-0.99$ is single-valued, as shown in Figure \ref{Figure3}.
\\We remind the reader that when the derivative
\begin{equation}
    \partial_{r_0}L_{app}(r_0)<0,\label{criterioestable}
\end{equation}
the embedding is stable. This is purely based on energy-considerations, see \cite{Chatzis:2024dlt, Avramis:2006nv}. For values of the parameter $\hat{\nu}\approx -1^+$, the stability criterion in eq.(\ref{criterioestable}) is not satisfied, characteristically indicating a transition. For values of $\hat\nu$ away from the lower edge of the allowed window, eq.(\ref{criterioestable}) is satisfied.
These qualitatively different behaviours have physical implications, as we explain below.

\begin{figure}[t]
\centering
\begin{subfigure}{0.5\linewidth}
\includegraphics[width=\linewidth]{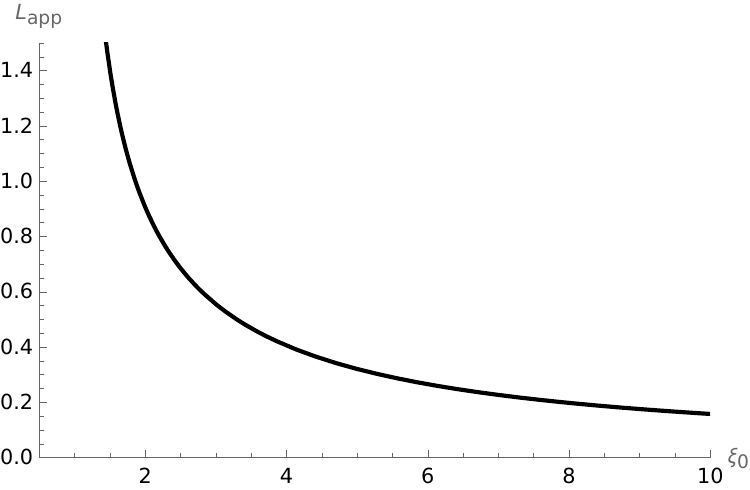}
\caption{Plot of the approximate length $L_{app}$ for  $\hat{\nu}=-0.99$ and $\theta_0=\frac{\pi}{2}$.}
\label{figuraLappi2}
\end{subfigure}
\hfill
\caption{Plot of the approximate separation of the quark-anti quark pair in eq.(\ref{Lapproxi}) for $\theta_0=\frac{\pi}{2}$.}
\label{Figure3}
\end{figure}
\begin{figure}[t]
\centering
\begin{subfigure}{0.5\linewidth}
\includegraphics[width=\linewidth]{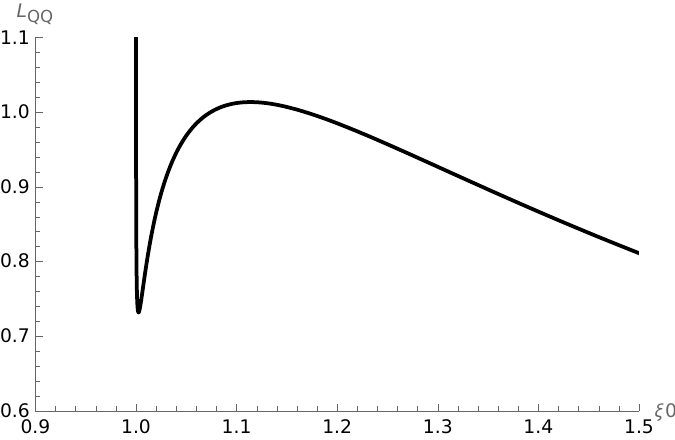}
\caption{Plot of the length $L_{QQ}(\xi_0)$ for  $\hat{\nu}=-0.99$.}
\label{Lxi0nu=-099}
\end{subfigure}
\hfill
\begin{subfigure}{0.44\linewidth}
\includegraphics[width=\linewidth]{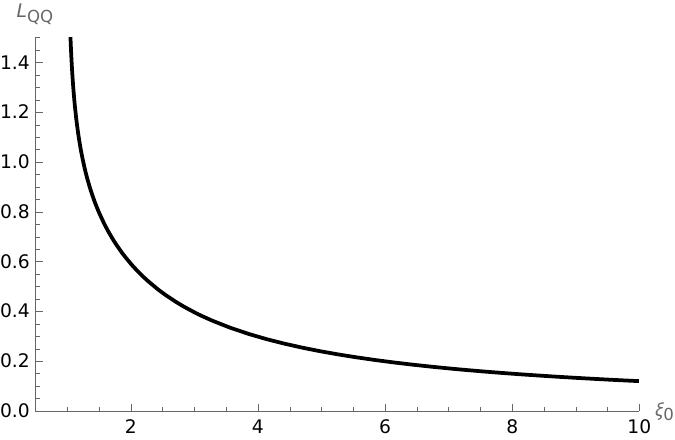}
\caption{Plot of the length $L_{QQ}(\xi_0)$ for  $\hat{\nu}=1$.}\label{Lxi0nu=1}
\end{subfigure}
\caption{Plots of $L_{QQ}$ in eq.(\ref{LQQexact}) for the two values $\hat\nu=-0.99$ and $\hat\nu=1$, {both for the same value of $\theta_0=0$}.}
\label{figure4}
\end{figure}
\begin{figure}[t]
\centering
\begin{subfigure}{0.5\linewidth}
\includegraphics[width=\linewidth]{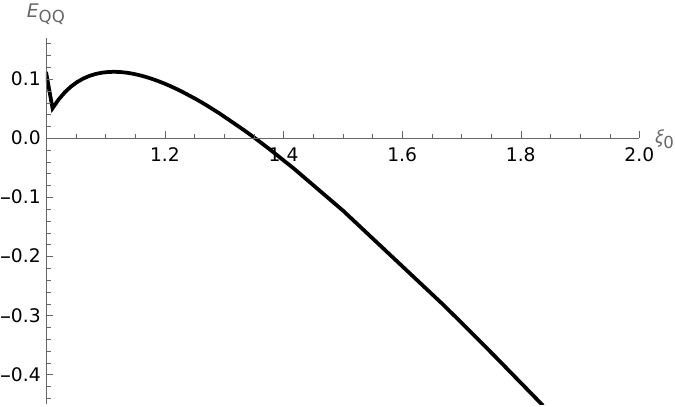}
\caption{Plot of $E_{QQ}(\xi_0)$ for  $\hat{\nu}=-0.99$, for $\theta_0=0$.}
\label{Exi0nu-099}
\end{subfigure}
\hfill
\begin{subfigure}{0.44\linewidth}
\includegraphics[width=\linewidth]{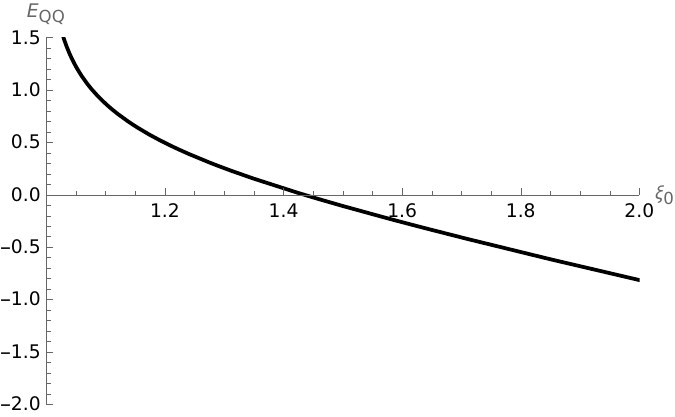}
\caption{Plot of  $E_{QQ}(\xi_0)$ for  $\hat{\nu}=1$, for $\theta_0=0$.}\label{Exi0nu1}
\end{subfigure}
\caption{Plots of the energy $E_{QQ}$ in eq.(\ref{EQQexact}) for the two values $\hat\nu=-0.99$ and $\hat\nu=1$.}
\label{figure5}
\end{figure}
\begin{figure}[t]
\centering
\begin{subfigure}{0.55\linewidth}
\includegraphics[width=\linewidth]{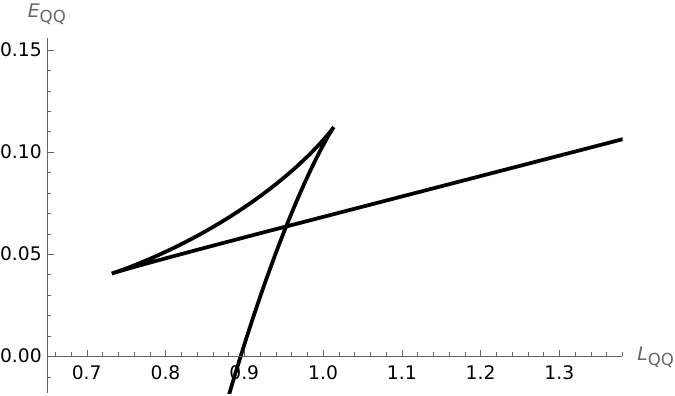}
\caption{Plot of $E_{QQ}(\xi_0)$ in terms of $L_{QQ}(\xi_0)$ for  $\hat{\nu}=-0.99$.}
\label{E-L-nu-099}
\end{subfigure}
\hfill
\begin{subfigure}{0.44\linewidth}
\includegraphics[width=\linewidth]{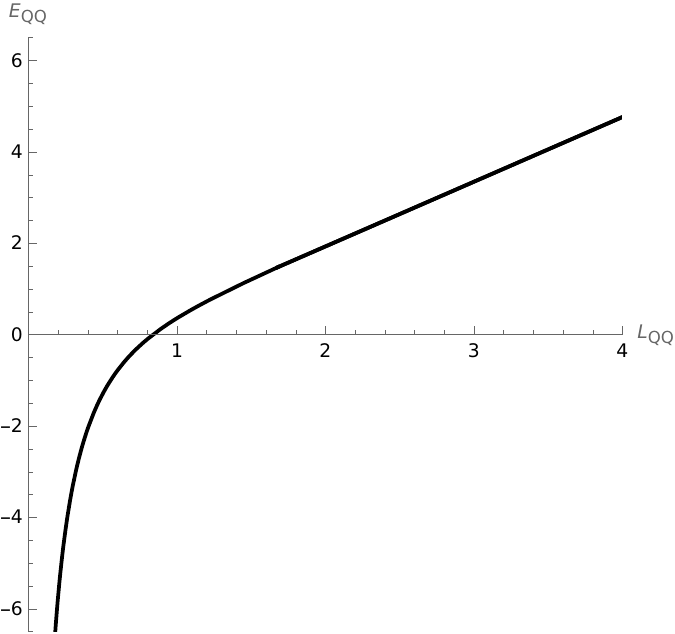}
\caption{Plot of  $E_{QQ}(\xi_0)$ in terms of $L_{QQ}(\xi_0)$ for the parameter  $\hat{\nu}=1$.}\label{E-L-nu1}
\end{subfigure}
\caption{Parametric plots of the energy $E_{QQ}$ in terms of $L_{QQ}$ for the two values $\hat\nu=-0.99$ and $\hat\nu=1$. We have set $\theta_0=0$}
\label{figure6}
\end{figure}

Let us now study the exact expressions for $L_{QQ}, E_{QQ}$ in eq.(\ref{WLintegrals}). As above, we consider the case where $q_1=q_2=Q$, and use the dimensionless variable $\xi=\frac{r}{r_\star}$, with parameters $\hat{\nu}=\frac{\varepsilon\ell^2}{r_\star^2}$ and $\hat{\mu}=\left(\frac{r_\star}{L}\right)^4$. We find
\begin{equation}
    \mathcal{F}^2 = \hat{\mu}\xi^2(\xi^2+\hat{\nu}\cos^2\theta_0),\quad \mathcal{G}^2 = \frac{\xi^4(\xi^2+\hat{\nu}\cos^2\theta_0)}{\xi^6+\hat{\nu}\xi^4-(1+\hat{\nu})}.
\end{equation}
From these, the length $L_{QQ}$ (the separation between the quark-antiquark pair) is written as an integral in terms of $\xi_0=\frac{r_0}{r_\star}$. It  reads,
\begin{eqnarray}
        L_{\mathrm{QQ}}(\xi_0)&=& 
        2 \, r_\star \xi_0\sqrt{\hat{\mu}^{-1}\xi_0^2 + \hat{\mu}^{-1}\hat{\nu}\cos^2\theta_0}  \, \times 
        \\[5pt]
        && \qquad \qquad  \qquad \int_{\xi_0}^{\infty}\frac{\mathrm{d}\xi~ \xi}{\sqrt{\left[\xi^6 + \hat{\nu}\xi^4 -(1+\hat{\nu})\right]\left[\xi^4-\xi_0^4+\hat{\nu}\cos^2\theta_0 (\xi^2-\xi_0^2)\right]}} \, . 
        \nonumber
    \label{LQQexact}
\end{eqnarray}
We plot this quantity for the same  values of $\theta_0=0$, $\hat{\mu}$ as above and for the two representative values of $\hat\nu$ chosen above. The result is found in Figures \ref{Lxi0nu=-099} and \ref{Lxi0nu=1}. Note the qualitative difference, one of the plots is single valued the other multivalued. Compare these plots with the approximate expression $L_{app}(\xi_0)$ displayed in Figures \ref{figuraLappa} and \ref{figuraLappb}.
Following the usual expressions in \cite{Nunez:2009da} we write an expression for the energy of the quark-anti quark pair, in terms of the turning point $\xi_0$. This reads,
\begin{eqnarray}
        E_{\mathrm{QQ}}(\xi_0)&=&\xi_0\sqrt{\hat{\mu}\xi_0^2+\hat{\mu}\hat{\nu}\cos^2\theta_0}L_{\mathrm{QQ}}(\xi_0)+2r_\star \int_{\xi_0}^{\infty}\mathrm{d}\xi\frac{\xi\sqrt{\xi^4-\xi_0^4+\hat{\nu}\cos^2\theta_0(\xi^2-\xi_0^2)}}{\sqrt{\xi^6+\hat{\nu}\xi^4-(1+\hat{\nu})}}
        \nonumber \\[5pt]
        &&-2r_\star \int_1^{\infty}\mathrm{d}\xi\frac{\xi^2\sqrt{\xi^2+\hat{\nu}\cos^2\theta_0}}{\sqrt{\xi^6+\hat{\nu}\xi^4-(1+\hat{\nu})}}.
    \label{EQQexact}
\end{eqnarray}
The resulting plots are depicted in Figures \ref{Exi0nu-099} and \ref{Exi0nu1}. Again, note the qualitatively different behaviours.
\\
To clarify the different physical behaviours, it is illustrative to parametrically plot $E_{QQ}$ in terms of $L_{QQ}$. For the same values of the parameters $\theta_0=0,\hat\mu$ and for the two representative values of $\hat\nu=-0.99$ and $\hat\nu=1$ these are displayed in Figures \ref{E-L-nu-099} and \ref{E-L-nu1}.
These plots are illustrative as they show the quark antiquark potential, in terms of the separation (the QFT physical observable). In the case $\hat\nu=-0.99$, we observe the signature 'swallow tail' indicating a phase transition. The transition is first order and usually associated with the competition between two different solutions for the string probe. For $\hat\nu=1$, the behaviour is neatly that in a confining field theory, without any trace of a phase transition. Actually, the 'swallow-tail' disappears  around $\hat{\nu}\sim -0.95$ in agreement with the plots in Figures \ref{figurariemann} and \ref{riccisquared}. 
\\
We propose that the transition observed in the case $\hat\nu=-0.99$ and $\theta_0\sim 0$ is {\it not a faithful representation} of the dynamics of the system. In fact, the large corrections afflicting the supergravity approximation discussed around Figure \ref{riccisquared}, suggest that for values of $\hat\nu$ close to the edge of allowed values, makes the calculations for $\hat{\nu}\approx -1^+$ {\it and} $\theta_0=0$, not trustable. In other words, we believe that the phase transition in the Wilson loop is not an actual physically relevant situation.
This is interesting, as the background is not singular in the range $1\leq\xi<\infty$, but the large values of higher invariants  (for $\hat\nu$ close to the edge of allowed values) indicate that the background should be corrected, changing its form. Hence our calculations for values of the parameter $\hat{\nu}\approx -1^+$ are not trustworthy for the embedding that sets $\theta_0=0$. 

Let us express this in a different language. The Coulomb branch solution of \cite{Freedman:1999gk, Freedman:1999gp} is singular because there are some light states in the QFT that the supergravity fails to capture. We avoid the singularity with the twisted compactification (equivalently introducing a mass gap in the theory), but the light states are degenerate with the mass-gap of the QFT. Hence they are needed for a correct description of the dynamics. Failing to include these states produces a dynamics of the Wilson loop that is not in agreement with the confining behaviour expected. As the parameter $\hat\nu$ moves away from the edge the mass gap and the light states separate and their influence in the dynamics ceases.

The structure and behaviour observed above for the Wilson loops in class I backgrounds repeats for the same observable in classes II and III as we analyse in what follows. Note that the backgrounds in classes II and III, we are in a situation similar to $\theta_0=0$.

\subsubsection{The Wilson loop in backgrounds of class II and III}
First, we discuss the same universal Wilson loop for the case of backgrounds of class III. For backgrounds of class II, we need to calculate using a wrapped M2 brane.

We follow the same steps and compute the Wilson loop for the type $\mathrm{IIA}$ background of eq.\eqref{10dGM-2}. The situation is more involved due to the nontrivial $r$-dependence of the functions $\tilde{f},\tilde{f}_i,X$ defined in eq.\eqref{IIA_functions}. We consider the embedding\footnote{In all the calculations concerning these class III background we  denote the worldsheet coordinate as $\hat{\sigma}$ to avoid confusion with the $\sigma$ coordinate of eq.\eqref{10dGM-2}.}:
\begin{equation}
    t=\tau,\quad w=\hat{\sigma},\quad r=r(\hat{\sigma}),
\end{equation}
where the rest of the coordinates take constant values. We specifically choose to localise the embedding in the $(\sigma,\eta)$ plane by taking $\eta=\eta_\star\in\mathbb{Z}$ with $0<\eta_\star<P$ and $\sigma\to 0$. One virtue of taking this limit is that information about the dual quiver theory, encoded in the rank function--defined as $\mathcal{R}(\eta)=\dot{V}\Big|_{\sigma=0}$, is easily accessed. The induced metric for the probe string reads,
\begin{equation}
    \mathrm{d}s^2_{\text{ind}}= 4\tilde{f}_1^{3/2}\tilde{f}_5^{1/2}\tilde{f}\Omega(r)\left\{ -\mathrm{d}t^2 + \left[1 + \frac{r^{\prime 2}}{r^2F(r)\lambda^4(r)\Om(r)} \right]\mathrm{d}w^2\right\}.
\end{equation}
Here $\Omega(r)=\frac{r^2\lambda^2(r)}{L^2}$ and the Nambu-Goto action is
\begin{eqnarray}
& & \mathrm{S}_{\mathrm{F}1} = \frac{1}{2\pi}\int \mathrm{d}t\mathrm{d}w\sqrt{-\mathrm{det}g_{\text{ind}}}=\frac{\mathcal{T}}{2\pi}\int _{-L_{\mathrm{QQ}}/2}^{L_{\mathrm{QQ}}/2}\mathrm{d}w\sqrt{\mathcal{F}^2 + \mathcal{G}^2r^{\prime 2}},\nonumber\\
& & \mathcal{F}^2= 16\tilde{f}_1^3\tilde{f}_5\tilde{f}^2\Omega^2(r),\quad \mathcal{G}^2 =\frac{16\tilde{f}_1^3\tilde{f}_5\tilde{f}^2\Omega(r) }{r^2F(r)\lambda^4(r)}. \label{F_and_G_WL_IIA_before_limit}
\end{eqnarray}
Here $\tilde{f}_i\equiv\tilde{f}_i(\sigma\to 0,\eta_\star,r)$. We make use of the expressions for the potential function $V(\sigma,\eta)$ and its derivatives expanded in Fourier-Bessel series, see eq.(\ref{Vsigmaeta}) and  also use the limiting behaviours,
\begin{equation}
  \text{As}\,\, x\to 0:\quad  xK_1(x)\to1, \,\, K_0(x) \approx -\gamma_{\mathrm{EM}}+\log\frac{2}{x}+\mathcal{O}(x^2).
\end{equation}
In the region $\sigma=\epsilon$ with $\epsilon\to 0$ we have,
\begin{equation}
\lim_{\epsilon\to0}\dot{V}(\epsilon,\eta)=\mathcal{R}(\eta),~~~\lim_{\epsilon\to0}\dot{V}^{\prime}(\epsilon,\eta)=\mathcal{R}^{\prime}(\eta),~~\lim_{\epsilon\to0}\ddot{V}(\epsilon,\eta)=0,~~~V^{\prime\prime}(\epsilon,\eta)\approx \hat{M}(\epsilon). \label{limites}
\end{equation}
where the last expression diverges as $\log\frac{1}{\epsilon}$ when $\epsilon\to0$. We can then figure the limits of the various functions of the type $\mathrm{IIA}$ background that appear in the expression of the induced metric. These are, 
\begin{equation}\label{typeIIA_limits}
    \begin{split}
        &\tilde{\Delta}\approx 2\hat{M}(\epsilon)\mathcal{R}(\eta),\quad Z^3\to X^3,\quad \tilde{f}_1\to \mathcal{R}(\eta)\\
        &\tilde{f}\to 1,\quad \tilde{f}_1^3\tilde{f}_5\approx 2X\mathcal{R}(\eta)\hat{M}^{-1}(\epsilon)\to 0,\quad \tilde{f}_2\to X^{-2},\quad \tilde{f}_3\to0.
    \end{split}
\end{equation}
Then  the values of ${\cal F},{\cal G}$ in eq.\eqref{F_and_G_WL_IIA_before_limit} as $\epsilon\to 0$ are,
\begin{eqnarray}\label{F_and_G_limits_WLI_IIA}
  & &  \mathcal{F}^2=\frac{32\mathcal{R}(\eta_\star)}{\hat{M}(\epsilon)}X(r)\Omega^2(r)= \frac{32\mathcal{R}(\eta_\star)}{\hat{M}(\epsilon)}\frac{r^4\lambda^6(r)}{L^4},\nonumber\\
    & &\mathcal{G}^2= \frac{32\mathcal{R}(\eta_\star) }{\hat{M}(\epsilon)}\frac{X(r)\Omega(r)}{r^2F(r)\lambda^4(r)}=\frac{32\mathcal{R}(\eta_\star)}{\hat{M}(\epsilon)}\frac{1}{F(r)}\;.
\end{eqnarray}
The dynamical $r$-dependent part is exactly the same as the functions appearing in embedding for the type $\mathrm{IIB}$ Wilson loop calculation. Indeed,  substituting the expressions $\Omega(r) = \frac{r^2\lambda^2(r)}{L^2}$ and $X(r)=\lambda^2(r)$, we compare with eq. \eqref{F_and_G_WL_IIB_embeddingI} after setting $\theta_0=0$. This implies that the analysis for the length and energy is the same up to the factor $\sqrt{\frac{32\mathcal{R}(\eta_\star)}{\hat{M}(\epsilon)}}$. This global factor indicates that the Wilson loop contains the information for which gauge node is being computed (given by ${\cal R}(\eta_\star)$). It is less clear what the physical meaning of the (vanishing factor) $\hat{M}(\epsilon)^{-1/2}$ is. The dynamical part of the Wilson loops is the same and this is the reason why we referred to the Wilson loop as an 'universal' observable.
The results are (up to a global numerical factor) those indicated in Figures \ref{Figure3}, \ref{figure4}, \ref{figure5}, \ref{figure6}.

There is one subtle difference with respect to the case of the Type IIB backgrounds. The backgrounds of class II and III are dual to field theories with fundamental degrees of freedom (flavours). Suppose that we allow the F1-string to probe the $\eta$-direction (instead of being fixed at $\eta_\star$ as we do here), by setting $\eta=\eta(\hat\sigma)$. The problem is more involved, depends on two dynamical variables $r(\hat{\sigma}),\eta(\hat\sigma)$ and the differential equations are more involved to solve and should be treated numerically. One example of this numerical treatment (for a system similar to the class III backgrounds) is found in \cite{Giliberti:2024eii}. One expects a behaviour similar to that observed in \cite{Giliberti:2024eii}, namely, the fundamental string to probe the $\eta$-space ending on a flavour group.

Let us now study the case of backgrounds in class II. For this, it is useful to consider a situation in which the sub-space $(v_1,v_2)$--the Riemann surface with metric $(dv_1^2+dv_2^2)$, is written as $(d\rho^2+ \rho^2 d\beta^2)$. We then consider an M2 brane, whose dynamics is parameterised in terms of three coordinates $(\sigma_1,\sigma_2,\sigma_3)$. The configuration we choose is
\begin{equation}
\sigma_1=t,~~\sigma_2=w, ~~\sigma_3=\beta, ~~~ r=r(w).    
\end{equation}
Using eq.(\ref{ds11_LLM}), we find the induced metric on the membrane is
\begin{eqnarray}
& & ds_{ind,M2}^2= G_{tt}dt^2 + (G_{ww} + G_{rr}r'^2)dw^2+ G_{\beta\beta}d\beta^2.\nonumber\\
& & G_{tt}= -e^{2\hat{\lambda}} {\cal A}^{1/3} \frac{4}{X} \frac{r^2\lambda^2(r)}{L^2}, ~~G_{ww}=-G_{tt},~~G_{rr}=e^{2\hat{\lambda}} {\cal A}^{1/3} \frac{4}{X} \frac{1}{r^2\lambda^4(r) F(r)},\nonumber\\
& & G_{\beta\beta}=-e^{2\hat{\lambda}} {\cal A}^{1/3} \frac{\partial_y e^{D_0}}{y}\rho^2,~~\text{where}~~{\cal A}= 1+ y^2 e^{-6\hat{\lambda}} (X^3-1).
\end{eqnarray}
The action of the membrane is
\begin{eqnarray}
& & S_{M2}\propto T_{M2}\int d^3\sigma \sqrt{-\det [g_{ab}]}.\nonumber\\
& & -\det[g_{ab}]=\frac{16\rho^2}{X^2} e^{6\hat{\lambda}}{\cal A}\left( -\frac{ \partial_y e^{D_0}}{ y}\right)\left(\frac{r^4 \lambda^4(r)}{L^4} + \frac{r'^2}{L^2 \lambda^2(r) F(r)} \right).
\end{eqnarray}
We should place the M2 brane at a position in $y$ such that $y\partial_y D_0\to\infty$. For example, for the function
$D_0(y,v_1,v_2)$, defined as $e^{D_0}=\frac{4(N^2-y^2)}{(1-v_1^2-v_2^2)^2}$ that solves eq.(\ref{todaeq}), we should place the probe M2 at $y=N$. In this case the prefactor of the determinant is $e^{6\hat{\lambda}}{\cal A}\frac{4 \partial_y e^{D_0}}{X^2 y}\sim X$, aside from sitting on a particular value of $(v_1,v_2)$ that we indicate as $\rho_\star$. The induced action on the membrane is of the form
\begin{eqnarray}
 S_{M2}\propto \left(T_{M2}\int d\beta \int dt\right)\int dw \sqrt{\frac{r^4\lambda^6(r)}{L^4} + \frac{r'^2}{L^2 F(r)}}.   
\end{eqnarray}
The dynamical part of this action is the same as that in eqs.(\ref{F_and_G_WL_IIB_embeddingI}),(\ref{F_and_G_WL_IIA_before_limit}). The dynamics of the membrane is the same as the dynamics of the probe string in backgrounds I and III.

\subsection{Flow central charge}\label{flowcentralsection}

A quantity of interest in all CFTs is the central charge, usually associated with its number of degrees of freedom. When we work with systems that flow away from a conformal point, a useful (but not strictly well-defined concept) is that of a central function. The idea is to write a quantity that at high energies reaches the value of the CFT central charge, whilst monotonically decreasing along the flow. If the IR QFT is gapped, we expect this central function to vanish in the far IR. In the case of the field theories dual to our backgrounds of type I, II and III, we clearly start from a fixed point, as the backgrounds all asymptote to AdS$_5$ in the UV (for $r\to\infty$). This fixed point is deformed by a VEV, as discussed in Section \ref{aspectsQFT}. The QFT is compactified (with a twist) along the $\phi$-direction. Hence, the system makes a transition from a four-dimensional CFT to a three dimensional gapped QFT. Writing a central function in this case seems hard using purely field theoretical means and quantities. In spite of this, holography provides us with a quantity, that is a direct generalization of the usual holographic central charge, see \cite{Macpherson:2014eza} for a definition at conformal points. The holographic generalization was written in \cite{Bea:2015fja} and more generally in \cite{Merrikin:2022yho}. Sometimes this is referred as the anisotropic central charge, as the backgrounds I, II and III can be thought as dual to anisotropic QFTs. In fact, note that the $SO(1,3)$ of the backgrounds in the UV is broken to $SO(1,2)\times SO(2)$ in the IR.

For an anisotropic QFT dual to a background  (and dilaton $\Phi$) of the form
\begin{equation}
    \mathrm{d} s^2=-a_0 \mathrm{d} t_0^2+\sum_{i=1}^{d} a_i \mathrm{d} x_i^2+\prod_{i=1}^{d}\left(a_i \right)^{\frac{1}{d}} \tilde{b} \mathrm{d} r^2+h_{a b} (\mathrm{d}y^a-A^a) (\mathrm{d}y^b-A^b), ~~~\Phi(\vec{y}, r).\label{backaniso}
\end{equation}
We compute the quantities
\begin{align}
& G_{i j} \mathrm{d} X^i \mathrm{d} X^j=\sum_{i=1}^{d} a_i \mathrm{d} x_i^2+h_{a b} (\mathrm{d} y^a-A^a) (\mathrm{d} y^b-A^b) \\
& \hat{H}=\left(\int \prod_{a=1}^{8} \mathrm{d} X^a \sqrt{e^{-4 \Phi} \operatorname{det}\left(G_{i j}\right)}\right)^2, \quad \frac{c_{\mathrm{flow}}}{\text{Vol}_d}  =d^d \frac{\tilde{b}^{\frac{d}{2}} \widehat{H}^{\frac{2 d+1}{2}}}{G_N^{(10)}\left(\widehat{H}^{\prime}\right)^d}.
\end{align}
In what follows we compute this quantity for each of the backgrounds I, II and III.
%


\subsubsection{Flow central charge for backgrounds of class I}

\begin{figure}[H]
    \centering
    \includegraphics[width=0.8\linewidth]{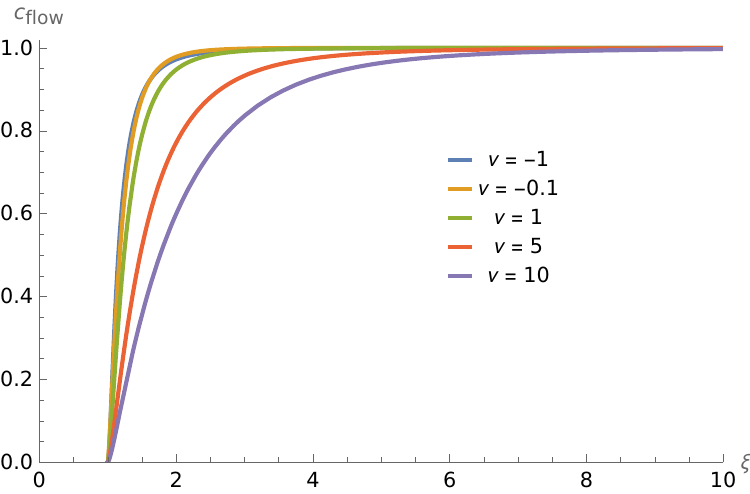}
    \caption{Plot of the conveniently normalised flow central charge in eq.(\ref{chols5}) as a function of $\xi$ for different values of $\hat{\nu}$ and $L=1$. }
    \label{Figure7}
\end{figure}

Comparing the background in eq.(\ref{S5-1}) with the generic one in eq.(\ref{backaniso}), we find
\begin{eqnarray}
& &
    \tilde{b} = \frac{L^{\frac{4}{3}}}{F(r)^{\frac{4}{3}}\lambda(r)^{6}r^4},~~~~~d=3, \label{btilde}\nonumber\\
& &V_{\text{int,I}}=\int \mathrm{d}^8x \; \sqrt{e^{-4\Phi}\operatorname{det}g_8} = L^3 r^3 \lambda(r)^3\sqrt{F(r)}\times\nonumber\\
& &\int \mathrm{d}\theta\! \cos^3\theta\! \sin\theta\! \int\! \mathrm{d}\psi \!\cos\psi\! \sin\psi \!\int\! \mathrm{d}\phi_1 \mathrm{d}\phi_2 \mathrm{d}\phi_3\mathrm{d}z\mathrm{d}w\mathrm{d}\phi \nonumber\\
& &     V_\mathrm{{int,I}} = \hat{\cal N}_{I} r^3 \lambda(r)^3 \sqrt{F(r)}, ~~
    \hat{\cal N}_{I} = L^3 \int \mathrm{d}\theta \cos^3\theta \sin\theta \int \mathrm{d}\psi \cos\psi \sin\psi \int \mathrm{d}\phi_1 \mathrm{d}\phi_2 \mathrm{d}\phi_3\mathrm{d}z\mathrm{d}w\mathrm{d}\phi,\nonumber\\
    & &\hat{H}= V_\mathrm{{int}}^2, ~~\frac{c_\mathrm{{flow}}}{\text{Vol}[z,w]L_\phi} = \frac{3^3 \;\tilde{b}^\frac{3}{2} \hat{H}^{\frac{7}{2}}}{G_{N}^{(10)}(\hat{H}')^3}.\nonumber   
\end{eqnarray}
Performing explicitly the calculation we find
\begin{eqnarray}
\frac{8G_N}{\hat{\cal N}_I L^2}  \frac{c_\mathrm{flow}}{\text{Vol}[z,w] L_\phi} &=& \!\left( \frac{\sqrt{F(r)}}{\lambda(r)\left[ \lambda(r) F(r) + r \lambda^\prime(r)F(r) + \frac{r\lambda(r) F^\prime(r)}{6} \right]} \right)^3 
\nonumber\\
 & = & \frac{27\; \xi^6}{(2\hat{\nu}+3\xi^2)^3} 
 \left(1+\frac{\hat{\nu}}{\xi^2}\right)^2\left(\frac{\xi^6-1+\hat{\nu}(\xi^4-1)}{\xi^4(\hat{\nu}+\xi^2)}\right)^\frac{3}{2} \, .\label{chols5}
\end{eqnarray}
We  plot this normalised quantity for different values of $\hat{\nu}$ (setting $L=1$) in Figure \ref{Figure7}.

We observe the following welcomed qualitative behaviours: the function is monotonic, it reaches the value of the CFT c-function in the UV, for large values of the $\xi$-coordinate. It vanishes in the far IR, for $\xi\approx 1$, indicating a gapped system.
Note that for this observable (in contrast with the Wilson loop), the needed corrections to the background close to the IR (for $\hat\nu\to -1^{+}$) do not seem to generate any unphysical behaviour.
Let us calculate the same observable in the case of the backgrounds II and III.


\subsubsection{Flow central charge for backgrounds of class II and III}
The procedure to calculate the flow central charge in these backgrounds is very similar to the one discussed above. The main difference being that now we have infinite families of backgrounds in classes II and III. 
For both classes II and III, we have the same $\tilde{b}(r)$ as in eq. (\ref{btilde}). For the quantity $V_{int}$
(note that class II backgrounds we compute a 9-volume and there is no dilaton) we find in each case,

\begin{eqnarray}
  V_\mathrm{int,II} &=& \hat{\cal N}_{II} \; r^3 \lambda(r)^3 \sqrt{F(r)},~~~ V_\mathrm{int,III} = \hat{\cal N}_{III} \; r^3 \lambda(r)^3 \sqrt{F(r)},\\
    \hat{\cal N}_\mathrm{II} &=& \frac{32}{L^2} \int \mathrm{d}\Omega_2  \mathrm{d}\chi    \int \mathrm{d}y \; \mathrm{d}v_1  \mathrm{d}v_2 \; y \partial_y e^{D_0}~\int \mathrm{d}z\mathrm{d}w\mathrm{d}\phi,\label{VintII}\\
 \hat{\cal N}_{III} &=&\frac{32}{L^2} \int \mathrm{d}\sigma \mathrm{d}\eta \; \dot{V}V^{\prime\prime}\sigma \;\int \mathrm{d}\Omega_2 ~\mathrm{d}\chi~\int  \mathrm{d}z\mathrm{d}w\mathrm{d}\phi
\end{eqnarray}
As a consequence, the quantity $c_{flow}$
normalised by the volumes of the $[z,w,\phi]$ is
\begin{eqnarray}
& &  \frac{8G_N}{\hat{\cal N}_{II} L^2}  \frac{c_\mathrm{flow,II}}{\text{Vol}[z,w] L_\phi} = 
\frac{8G_N}{\hat{\cal N}_{III} L^2}  \frac{c_\mathrm{flow,III}}{\text{Vol}[z,w] L_\phi} =\frac{8G_N}{\hat{\cal N}_{I} L^2}  \frac{c_\mathrm{flow,I}}{\text{Vol}[z,w] L_\phi}.
\end{eqnarray}
The plot of the (conveniently normalised) flow central functions for the backgrounds of class II and III is the same as in Figure \ref{Figure7}. Note that this (like the Wilson loop) is a universal observable quantity. Importantly, note that the quantities $\hat{\cal N}_{II}$ and $\hat{\cal N}_{III}$ are the central charges (free energies) of the UV CFTs corresponding to the Gaiotto CFTs. For each different function $D_0(y,v_1,v_2)$ and $V(\sigma,\eta)$ we  have a dual to  a different  ${\cal N}=2$ SCFT$_4$ with its own central charge characterising the field theory. Indeed, for the UV-CFTs the central charge (a number) can be calculated to be proportional to $\tilde{\cal N}_{II}$ and ${\tilde{\cal N}}_{III}$. This result is obtained either by holographic means \cite{Lozano:2016kum,Nunez:2019gbg,Macpherson:2024frt} or by a localisation-matrix model calculation \cite{Nunez:2023loo}.
\\
Let us now study a different quantity of interest.
\subsection{Complexity: a puzzle and its solution}
There are various conjectural definitions of complexity. In this work we adopt the complexity-equals-volume one.
In this context, one specifies some spatial slice $\Sigma$ on the boundary of the spacetime. The complexity $\mathcal{C}$ of a pure state $|\Psi\rangle$ of a holographic field theory on this patch will be given by the volume of a co-dimension one slice B in the bulk. This sub-manifold in the bulk, whose volume obeys some maximal condition, has its boundary on $\Sigma$. In particular, it was proposed in \cite{Fatemiabhari:2024aua} that the complexity is calculated as the  (conveniently normalised by the overall warp factor) nine-volume (in eleven dimensional backgrounds is a ten-manifold) defined by the coordinates
$[w,z,\phi,r,{\cal M}]$. Here $\cal M$ is the deformed $S^5$ in the backgrounds of class I. For class II solutions is defined by the six coordinates $[y,\Omega_2,\chi,v_1,v_2]$. Finally for class III backgrounds $\cal M$ is the five manifold defined by $[\sigma,\eta,\Omega_2,\chi]$. As proposed in \cite{Fatemiabhari:2024aua}, we calculate in the ten dimensional cases I and III,
\begin{equation}
    \mathcal{C}_{V} \propto \frac{1}{G_N^{(10)}}\int \; \mathrm{d}^9x \sqrt{\frac{e^{-4\Phi}\operatorname{det}g_9}{\mathcal{A}}}.\label{complexity10}
\end{equation}
Whilst for the eleven dimensional backgrounds of class II we have,
\begin{equation}
    \mathcal{C}_{V} \propto \frac{1}{G_N^{(11)}}\int \; \mathrm{d}^{10}x \sqrt{\frac{\operatorname{det}g_{10}}{\mathcal{A}}}.\label{complexity11}
\end{equation}
Below, we encounter a puzzling result: the complexity of backgrounds in classes II and III is different from that in backgrounds of class I. This is odd. Similarly to the quantities in Sections \ref{wilsonsection} and \ref{flowcentralsection}, we expect a universal result, with the same structure as the one outlined, namely a factor corresponding to the UV-CFT$_4$ times a  dynamical factor, dependent on the flow. The resolution of this puzzle takes us to a more refined version of how the complexity of a QFT should be computed using the holographic dual. Let us first present the naive calculation.

For backgrounds I, II and III we have
\begin{eqnarray}
& &     \mathcal{A}_{I} = \frac{\zeta(r,\theta)}{L^2}, ~~~ \mathcal{A}_{II}= \frac{4 e^{2\hat\lambda}}{X} \left[1+y^2e^{-6\hat\lambda}(X^3-1)\right]^{1/3},~~ \mathcal{A}_{III} = 4\tilde{f}(\tilde{f}_1^3\tilde{f}_5)^{\frac{1}{2}}.\label{calAI-II-III}
\end{eqnarray}
Calculating according with eqs.(\ref{complexity10})-(\ref{complexity11}) we have,
\begin{equation}
    \mathcal{C}_{V,I} \propto \Bigg[\frac{L^3}{G_N^{(10)}}\int \; \mathrm{d}\theta \; \cos^3\theta \sin\theta \int \mathrm{d}\psi \cos\psi\sin\psi \int \;\mathrm{d}w\;\mathrm{d}z\; \mathrm{d}\phi~\mathrm{d}\phi_1\;\mathrm{d}\phi_2 \;\mathrm{d}\phi_3\Bigg] \times \int_{r_\star}^{\Lambda}  \mathrm{d}r ~r^2.\label{complexityI}
\end{equation}
For backgrounds in class II, we find
\begin{equation}
    \mathcal{C}_{V,II} \propto \Bigg[\frac{2^{5}}{G_N^{(11)}L^2} \int \mathrm{d}y ~y ~\partial_y e^{D_0}\int  \; \mathrm{d}w~ \mathrm{d}z\; \mathrm{d}\phi \; \mathrm{d}\chi \; \mathrm{d}v_1 \mathrm{d} v_2\; \mathrm{d}\Omega_2\Bigg] \times \int_{r_\star}^{\Lambda} \mathrm{d}r \; r^2 \lambda(r)
    . \label{complexityII}
\end{equation}
Finally, for backgrounds in class III we find,
\begin{equation}
\mathcal{C}_{V,III} \propto
   \Bigg[ \frac{2^5 }{G_N^{(10)} L^2} \int \mathrm{d}\sigma~ \mathrm{d}\eta \; \dot{V}~V^{\prime\prime}~ \sigma \; \int\mathrm{d}w \; \mathrm{d}z \; \mathrm{d}\phi  \mathrm{d}\Omega_2 d\chi\Bigg]\times \int_{r_\star}^{\Lambda} \mathrm{d}r \; r^2 \lambda(r). \label{complexityIII}
\end{equation}
Note that in the backgrounds of classes II and III, the factor of $\lambda(r)$ appears in the respective integration over the radial coordinate, eqs.(\ref{complexityII}),(\ref{complexityIII}). This is in contrast with the result for the backgrounds of class I, in eq.(\ref{complexityI}).
This is a puzzling result. Given the previous 'universal behaviours' observed for the quantities in Sections \ref{wilsonsection} and \ref{flowcentralsection}, why is the complexity failing to behave in the same way?

The resolution to the puzzle starts by recognising that the string-frame metric for the Type IIB background of class I can be written as,
\begin{eqnarray}
& & ds_{10,st}^2=\Delta^{1/2}\Bigg[\frac{r^2 \lambda^2(r)}{L^2}\left(-dt^2+ dz^2+dw^2+ L^2 F(r) d\phi^2 \right) +\frac{dr^2}{r^2\lambda^4(r) F(r)} +\label{otramanera}\\
& & \frac{L^2}{\Delta \lambda^2(r)}\left[d\theta^2\zeta^2(r,\theta) +\cos^2\theta d\psi^2 +\mu_1^2 (D\phi_1)^2 +\mu_2^2 (D\phi_2)^2 +\mu_3^2\lambda^6(r) (D\phi_3)^2\right]\Bigg],\nonumber\\   
& & \text{where}~~ \Delta= \frac{1}{\lambda^4(r)}(1+ \frac{\varepsilon \ell^2}{r^2}\cos^2\theta),~~\mu_1=\cos\theta\cos\psi,~~\mu_2=\cos\theta\sin\psi,~~\mu_3=\sin\theta.\nonumber
\end{eqnarray}
Calculating the nine-dimensional metric obtained setting $t=$constant in eq.(\ref{otramanera}) and using that  ${\cal A}=\Delta^{1/2}$ we find,
\begin{eqnarray}
    & & \mathcal{C}_{V,I} \propto \Bigg[\frac{L^3}{G_N^{(10)}}\int \; \mathrm{d}\theta \; \cos^3\theta \sin\theta \int \mathrm{d}\psi \cos\psi\sin\psi \int \;\mathrm{d}w\;\mathrm{d}z\; \mathrm{d}\phi~\mathrm{d}\phi_1\;\mathrm{d}\phi_2 \;\mathrm{d}\phi_3\Bigg] \times \nonumber\\
    & & ~~~~~~~~~\int_{r_\star}^{\Lambda}  \mathrm{d}r ~r^2 ~\lambda(r).\label{complexityIcorrect}
\end{eqnarray}
We then see the same result (we refer to the dynamical part involving the integral in the $r$-direction) as for the other two classes of backgrounds.

Why is this working like this? Or, why do we need to write the backgrounds of class I as in eq.(\ref{otramanera}) to compute the complexity? One way of understanding this is inspired by lower-dimensional gauged supergravity. 
As we explain in \cite{CHINZ2}--{see Section \ref{rough-view} for a summary}--
backgrounds of class I, II and III are obtained by lifting one single solution of five dimensional gauge supergravity, see \cite{Anabalon:2024che}. The complexity is defined in such a way that the factor ${\cal A}$ by which we quotient the determinant of the nine-manifold is the factor arising when we lift this gauged supergravity solution to 10 or 11 dimensions. This justifies also the presence of the factor  $\tilde{f}$ in ${\cal A}_{III}$.

Note that the expressions in between brackets in eqs.(\ref{complexityII})-(\ref{complexityIII}) and (\ref{complexityIcorrect}) are proportional to the UV-value of c$_{flow}$, namely $c_{flow}(\xi\to\infty)$, which in turn is related to  the central charge of the UV-CFT. Hence the complexity is proportional to the central charge of the UV-CFT.
Had we not included the factor of $\tilde{f}$ in ${\cal A}_{III}$, the result would have not been proportional to the central charge of the UV-CFT.


\section{Summary and conclusions}\label{conclusionsection}
In this paper, we elaborate on the recently discovered type IIB background of \cite{Anabalon:2024che}--here referred as class I, along with two new families of backgrounds presented here for the first time.
The first new family (class II) is a solution of $D=11$ supergravity, belonging to the Lin-Lunin-Maldacena/Gaiotto-Maldacena family of backgrounds in \cite{Lin:2004nb,Gaiotto:2009gz}, which are dual to 4d ${\cal N}=2$ non-Lagrangian SCFTs.
The second family of backgrounds (class III) is a type IIA solution obtained via dimensional reduction from the 11d background and lies within the linear quiver class \cite{Gaiotto:2009gz}.
In all three cases, the boundary CFT is deformed by the VEVs of two operators: a dimension-three operator that mixes part of the Lorentz group with the R-symmetry while preserving supersymmetry, and a dimension-two operator triggering a flow along the Coulomb branch \cite{Freedman:1999gk}.
These deformations generically preserve supersymmetry.

To analyze the backgrounds discussed in Section \ref{SUGRA_backgrounds} and their field theory duals, we compute a variety of observables holographically.
Some of our observables are chosen to probe only the field theory directions and remain insensitive to the specific structure of the internal space (five or six-dimensional). Other observables fully explore the internal space.
A notable outcome of these computations is the emergence of universality in the observables' behaviour.
Although this might be anticipated due to the choice of observables (as in similar holographic studies, e.g., \cite{Chatzis:2024kdu, Chatzis:2024top}), it is nonetheless remarkable that distinct CFTs exhibit such shared features.
It would be particularly interesting to extend the holographic analysis to probe some part of the internal space, potentially breaking the observed universality.

The Wilson loop calculations in all three classes reduce to expressions that naturally split into two parts: a dynamical, $r$-dependent term common to all cases, and a class-specific term characterizing the underlying gauge theory.
Additional differences include the presence of fundamental degrees of freedom in class II and III backgrounds (absent in class I), and the number of string trajectories studied, two in class I, one in classes II and III.
Nevertheless, the central conclusion is that the deformed QFT confines external non-dynamical quarks. This is valid in the case for backgrounds of classes II and III if we do not allow our F1 probe to explore the direction associated with the flavours of the CFT (or similarly, that the flavours are very energy-expensive to produce). 
In the class I background, we identify a region in parameter space that may signal a first-order phase transition, though we argue this is likely due to the string approaching a region, where the supergravity approximation breaks down, even in the absence of a naked singularity.

The holographic central charge calculation exhibits the expected monotonic behaviour: flowing from the UV value of the CFT $c$-function to zero in the IR, reflecting a gapped QFT.
As in the Wilson loop case, the central charge expression factors into a UV contribution (capturing the CFT data) and a dynamical part common to all three backgrounds.

We also study the concept of holographic complexity, interpreted in the quantum circuit framework as the number of elementary gates needed to construct a given Hilbert space state.
We adopt the proposal of \cite{Fatemiabhari:2024aua}, in which the nine- or ten-dimensional volume (excluding time) is appropriately normalized with the warp factor.
To preserve the universal and factorized behaviour observed in the Wilson loop and central charge calculations, it is crucial that the warp factor be chosen according to the lower-dimensional gauged supergravity from which the  background originates after uplift.

To deepen our understanding of the dual field theories, a natural next step would be to introduce a probe brane and study its dynamics and fluctuations.
The meson spectrum of the field theory corresponds to the fluctuation modes of the probe brane on the gravity side, leading to a discrete and gapped spectrum and analogously the spectrum of glueballs following the formalism and developments in \cite{Elander:2025fpk, Fatemiabhari:2024lct}. The calculation of Entanglement Entropy, on spherical regions, or the time-like version could be done in these backgrounds, along the lines of the papers \cite{Jokela:2025cyz, Afrasiar:2024lsi, Nunez:2025gxq}. Another idea would be to study the fluctuations and spectrum of a F1-probe. The presence of higher curvature invariants in such calculations \cite{Bigazzi:2024biz} might teach us about the states not included in supergravity. We discuss these and other aspects in future work.

\section*{Acknowledgments} We would like to thank very useful conversations with: Andr\'es Anabal\'on, Francesco Bigazzi, Federico Castellani, Aldo Cotrone, Mauro Giliberti, Prem Kumar, Horatiu Nastase, Marcelo Oyarzo, Dibakar Roychowdhury, Colin Sterckx, Ricardo Stuardo. The research of G. I. is supported by the Einstein Stiftung Berlin via the Einstein International Postdoctoral Fellowship program ``Generalised dualities and their holographic applications to condensed matter physics'' (project number IPF- 2020-604). D.C. has been supported by the STFC consolidated grant ST/Y509644-1. M.H. has been supported by the STFC consolidated grant ST/Y509644/1. C. N. is supported by the grants ST/Y509644-1, ST/X000648/1 and ST/T000813/1.

\bibliographystyle{utphys}
\bibliography{refs}

\end{document}